\documentclass{jfm}
\usepackage{graphicx}
\usepackage{epstopdf, epsfig}
\usepackage{siunitx}
\usepackage{tablefootnote}
\usepackage{textgreek}
\usepackage{xcolor}
\usepackage{cleveref}
\usepackage{array}

\newcommand{\MassFraction}{c}
\newcommand{\VaporField}{c^\text{g}}
\newcommand{\VaporFieldVLE}{c^\text{g,VLE}}

\newcommand{\SurfTension}{\gamma}
\newcommand{\MarangoniNumber}{\mathrm{Ma}}
\newcommand{\SchmidtNumber}{\mathrm{Sc}}

\newcommand{\VaporDiffusivity}{D^\text{g}}

\newcommand{\VeloX}{u_x}
\newcommand{\VeloZ}{u_z}
\newcommand{\MassFractEth}{\MassFraction_\text{e}}
\newcommand{\GasDensity}{\rho^\text{g}}
\newcommand{\SatPressure}[1]{p_{#1,\text{sat}}}
\newcommand{\MolarMass}{M}
\newcommand{\MoleFract}{n}
\newcommand{\ActivityCoeff}{\mathcal{A}} 
\newcommand{\DerSurfTenMassFracDim}{d\SurfTension_0^\mathrm{d}/d\MassFractEth}
\newcommand{\DerSurfTenMassFracNonDim}{d\SurfTension/d\MassFractEth}

\shorttitle{Marangoni instability triggered by selective evaporation}
\shortauthor{R. A. Lopez de la Cruz, C. Diddens, X. Zhang and D. Lohse}

\title{Marangoni instability triggered by selective evaporation of a binary liquid inside a Hele-Shaw cell}

\author{Ricardo Arturo Lopez de la Cruz\aff{1}
	\corresp{\email{r.a.lopezdelacruz@utwente.nl}},
	Christian Diddens\aff{1,2}
	\corresp{\email{c.diddens@utwente.nl}},
  Xuehua Zhang\aff{3,1}
  \corresp{\email{xuehua.zhang@ualberta.ca}}
 \and Detlef Lohse\aff{1,4}
 \corresp{\email{d.lohse@utwente.nl}}}

\affiliation{\aff{1} Physics of Fluids Group, Max-Planck-Center Twente for Complex Fluid Dynamics, Mesa+ Institute, and J. M. Burgers Centre for Fluid Dynamics, Faculty of Science and Technology, University of Twente, P.O. Box 217, 7500 AE Enschede, The Netherlands
\aff{2} Department of Mechanical Engineering, Eindhoven University of Technology, P.O. Box 513, 5600 MB Eindhoven, The Netherlands
\aff{3}Department of Chemical and Materials Engineering, University of Alberta, Edmonton, Alberta, T6G 1H9, Canada
\aff{4} Max Planck Institute for Dynamics and Self-Organization, Am Fa\ss berg 17, 37077 G\"{o}ttingen, Germany}

\begin{document}

\maketitle

\begin{abstract}

Interfacial stability is important for many processes involving heat and mass transfer across two immiscible phases. When this transfer takes place in the form of evaporation of a binary solution with one component being more volatile than the other, gradients in surface tension can arise. These gradients can ultimately destabilise the liquid-gas interface. In the present work, we study the evaporation of an ethanol-water solution, for which ethanol has a larger volatility. The solution is contained in a horizontal Hele-Shaw cell which is open from one end to allow for evaporation into air. A Marangoni instability is then triggered at the liquid-air interface. We study the temporal evolution of this instability by observing the effects that it has on the bulk of the liquid. More specifically, the growth of convective cells is visualized with confocal microscopy and the velocity field close to the interface is measured with micro-particle-image-velocimetry. The results of numerical simulations based on quasi 2D equations satisfactorily compare with the experimental observations, even without consideration of evaporative cooling, although this cooling can play an extra role in experiments. Furthermore, a linear stability analysis applied to a simplified version of the quasi 2D equations showed reasonably good agreement with the results from simulations at early times, when the instability has just been triggered and no coarsening has taken place. In particular, we find a critical Marangoni number below which a regime of stability is predicted.

\end{abstract}


\section{Introduction}

The stability of an interface is important for processes where mass or heat transfer between two phases take place, e.g. drying of paint, coating, distillation, absorption, and extraction. In the particular case of a Marangoni instability, variations of temperature or concentration along the interface give rise to surface tension gradients that trigger convection. The study of interfacial instability goes back to the pioneering work of \citet{Benard_1901}. Although in that work natural convection was considered to be the cause of the instability, \citet{pearson_1958} showed by means of linear stability analysis that some of the observations could be explained by gradients in surface tension caused by variations in temperature (thermal Marangoni instability). Soon after, \citet{SternlingScriven1960AIChE} showed, by a linear stability analysis too, that the same effect could be driven by gradients of concentration (solutal Marangoni instability).

Since then, a large body of studies has extended and generalized the works by \citet{SternlingScriven1960AIChE}, for example, to three dimensions with deformable interfaces \citep{scriven_sternling_1964, Hennenberg_1977} and gravity waves \citep{smith_1966}. Also the effects of chemical reactions \citep{RUCKENSTEIN1964}, finite domains \citep{REICHENBACH1981433}, finite domains with binary mixtures and the Soret effect \citep{bergeon_1998}, and modified geometries (such as spherically symmetric \citep{Sorensen_1980}) have been considered. Exponential concentration profiles and surfactant accumulation at the interface were studied by \citet{Sorensen_1977,Sorensen_1978}. More recently, \citet{Bratsun2004} studied the instability between two liquids inside a Hele-Shaw cell where a chemical reaction takes place at the interface. \citet{picardo2016solutal} considered the case of evaporation coupled with a Poiseulle flow and \citet{MACHRAFI2010} looked at the stability of an evaporating ethanol-water mixture in two dimensions taking evaporative cooling and the Soret effect into account. This list is by no means extensive, the reviews by \citet{Levich_1969}, \citet{Sanfeld1984} and \citet{Kovalchuk2006marangoni} present a more comprehensive overview, including cases where heat transfer is also considered. We refer to the reviews of \citet{Linde_2013}, and \citet{Schwarzenberger_2014} for further experimental and numerical work.

While a considerable part of the experimental work has focused on extended interfaces \citep{Linde_2013,Zhang_Behringer_2011}, there has also been quite some work on binary systems confined within a Hele-Shaw cell. Employing this kind of cell has the advantage that it allows to have visual access to the effects that the instability has in the bulk of the two phases. Furthermore, the system can be considered as quasi 2D. Many of the papers that make use of a Hele-Shaw cell have focused on liquid-liquid interfaces where chemical reactions take place, including experimental \citep{Eckert_2004, Shi_2006, Shi_2007, Shi_2008, Eckert_2012, Schwarzenberger_2012}, numerical \citep{Mokbel_2017, grahn2006two}, and theoretical \citep{Bratsun2004} approaches. In particular, \citet{Kollner_2015} performed both experiments and numerical simulations, showing good qualitative agreement between them, though the Marangoni rolls in experiments grew faster than in simulations. 

Following the criteria established by \citet{SternlingScriven1960AIChE}, \citet{Linde_1964_HS} studied the stability of liquid-gas interfaces of binary systems, where either selective evaporation or adsorption triggered a Marangoni instability. \citet{Linde_1964_HS} pointed out that both thermal and solutal instabilities were at play simultaneously. By looking at the evaporation of pure liquids to isolate the thermal instability, they observed that the flow was weaker than when the solutal Marangoni instability was also present. In this way they were able to suggest that solutal Marangoni was the strongest mechanism. However, in that paper, the quantitative description of the phenomena was limited. 

More generally, great efforts have been devoted to the understanding of the physicochemical hydrodynamics of multicomponent systems \citep{Levich_1962physicochemical}, especially of multicomponent droplets, due to their importance in many fields like chemical diagnosis, injekt printing or nanotechnology, to mention a few. For a recent review on this subject we refer to \citet{Lohse_2020}. In many of these systems mass transfer through interfaces causes gradients in chemical concentration that in turn results in a Marangoni flow. Such a flow can then lead to quite rich phenomena, for example, it can make a droplet inside a bath jump against the pull of gravity \citep{Li_2019_Bouncing, li2021marangoni}, and more generally cause the self-propulsion of droplets \citep{izri2014self, maass_2016, Lohse_2020}. These Marangoni flows can also assist the emulsification of evaporating ternary droplets \citep{Tan_2016, Tan_2017}, or drive the so-called Marangoni bursting \citep{keiser2017marangoni}. In some cases Marangoni forces compete with gravity, leading to unexpected flow directions inside evaporating sessile and pendant droplets \citep{Li2019, diddens2020competing}. Clearly, these phenomena are closely connected with the presence of instabilities at interfaces where mass transfer and evaporative cooling take place. However, the three-dimensional shape of the droplet which changes in time makes the geometry complicated. Therefore, the Hele-Shaw geometry is a good candidate to further understand the Marangoni convection present when selective evaporation takes place. Morever, the very common use of microfluidic devices in research nowadays makes it important to understand the influence that the confinement has on the Marangoni convection. Indeed, Marangoni flow in microfluidic devices can be used to enhance mixing \citep{Michelin_2020}.

The aim of this work is to revisit one of the systems studied by \citet{Linde_1964_HS}, namely a binary solution of ethanol-water confined by a Hele-Shaw cell evaporating into air. In particular, we study the evolution of the instability in time by means of confocal microscopy and $\mu$-PIV (section 2). In section 3, this system is modelled and numerical simulations of the model equations are performed. This model is quasi-2D and takes into account the drag caused by the walls of the cell \citep{Kollner_2015, Bizon1997_Chaos, Bratsun2004}. The simulation results are able to reproduce our experimental observations quite well, even though we do not consider the effects of evaporative cooling. In section 4 we present a linear stability analysis of a simplified version of the quasi-2D model equations. In section 5 we compare its results with those from numerical simulations at early times, showing not perfect, but close agreement. Finally, in section 6 we conclude and give an outlook on future work with ternary systems.

\section{Experimental methods and results}

\begin{figure}
	\centerline{\includegraphics[width=\linewidth]{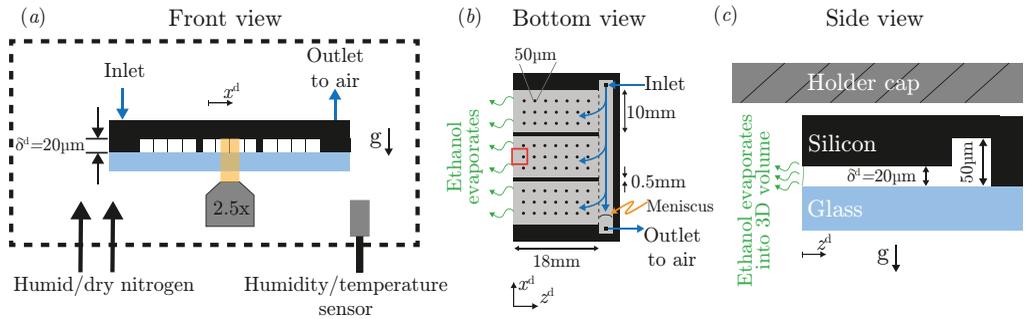}}
	\caption{Sketch of the experimental setup from (\textit{a}) frontal, (\textit{b}) bottom, and (\textit{c}) side views. The main component was a microfluidic chip made of an edged silicon wafer and a glass plate for observation. The gap between both plates was 20 \si{\micro\metre}. To maintain a homogeneous gap, a matrix of pillars was placed between the wafer and the glass, represented as thin vertical lines in (\textit{a}) and as circles in (\textit{b}) (not to scale). Additionally, two walls divided the cell in three different parts. At the back of the chip a channel of 50 \si{\micro\metre} in depth was used to continuously refill the cell and compensate for mass loses due to evaporation. As depicted in (\textit{a}), the whole chip was surrounded by a chamber where the humidity was monitored and controlled with a flow of dry and humid nitrogen. In (\textit{b}), the gap region is represented in grey and the red rectangle indicates the region of observation. In (\textit{c}), the meniscus is drawn (solid green line) to indicate a slightly wetting liquid. The dimensions are not to scale for better visualization. The silicon wafer is 400 \si{\micro\meter} thick without edging and the glass plate is 500 \si{\micro\meter} thick. The holder cap was about 170 \si{\micro\meter} above the chip.}
	\label{fig:SketchSetUp}
\end{figure}

\subsection{Materials and methods}
\subsubsection{Design of Hele-Shaw cell}
A microfluidic chip (Beijing First Mems Co., Ltd) was used as a Hele-Shaw cell. In figure \ref{fig:SketchSetUp}(\textit{a}) a cross section of the chip is shown. The black region was made of an edged silicon wafer, while the light blue plate was made of glass to allow for optical visualization. The gap between the two plates was of 20 \si{\micro\metre} and kept constant by means of a series of pillars depicted as thin vertical lines in figure \ref{fig:SketchSetUp}(\textit{a}). Further details can be seen in appendix \ref{app:FurtherExpDetails_Pillars}. 
At the back of the chip (right edge in figure \ref{fig:SketchSetUp}(\textit{b}) and (\textit{c})), a channel 50 \si{\micro\metre} in depth was used to fill the chip with the binary liquid. Three sides of the chip were closed and one was open to allow for evaporation (left side in figure \ref{fig:SketchSetUp}(\textit{b}) and (\textit{c})). The surface of the chip was made hydrophobic as described in appendix \ref{app:FurtherExpDetails_Coating}.

\subsubsection{Experimental setup}
We visualized the evaporation process at a region around the open edge (red square in figure \ref{fig:SketchSetUp}(\textit{b})) with a confocal microscope (Nikon Confocal Microscopes A1 system). As opposed to the liquid, the gas phase was not contained by the Hele-Shaw cell; instead it was a three-dimensional volume of about 3000 \si{\cubic\centi\meter}. This volume was created by covering part of the microscope with plastic wrap (see SI). The relative humidity (RH) in this chamber was maintained within a desired range using a negative feedback. Every time, the RH was outside the desired range, a flow of either humid or dry nitrogen was gently blown in. As long as the RH was within the range, the nitrogen flow was off. Humidity and temperature sensors (HIH6130, Honeywell) were used to monitor these two quantities. The readings were taken with an Arduino board (Arduino Nano) which also controlled the valves of the humid and dry nitrogen accordingly. The range over which there was no flow was either 50 $\pm$ 1\% or 51 $\pm$ 1\%, depending on the RH in the room. This resulted in an overall range between 48-52\% for all our experiments. Examples of the time series of the relative humidity in the chamber and extra details on the chamber can be found in the SI.

In the case of the temperature of the air in the chamber, we only measured it during each experiment, but did not control it. Due to the heating caused by the electronics, the temperature was constantly rising at about 2 \si{\kelvin\per\hour}. In the majority of our experiments the focus was on the first 100 \si{\second}, during which the variations in temperature stayed within 0.3 \si{K}. The initial temperature of different experiments ranged from 294 to 298 \si{K}, with the majority starting around 297 \si{K}. We did not observe a strong effect on the main evolution of the instability and the velocity field. Examples of the temperature time series can be found in the SI.

We assumed that at the beginning of each experiment the concentration of ethanol in the gas phase was negligible. We did not introduce ethanol into the chamber intentionally besides that coming from evaporation of the liquid inside the cell. Between each experiments, the chamber was opened to change the chip, allowing the ethanol that evaporated during the experiment to escape. 

In each experiment, the solution was delivered from a syringe connected through a capillary tube to the back channel of the chip (see figure \ref{fig:SketchSetUp}(\textit{b})). The flow rate was controlled with a motorized syringe pump (Harvard Instruments, PHD 2000). In each experiment the chip was filled within a minute at a rate of 15 \si{\micro\liter\per\minute}. After the chip was filled, we had a delay of about 30 \si{\second} to 60 \si{\second} before a replenishing flow was started to compensate the mass loss due to evaporation. This flow was kept constant throughout the rest of the experiment  (the direction of the flow is depicted with blue arrows in figure \ref{fig:SketchSetUp}(\textit{b})). Despite the hydrophobicity of the chip, the ethanol-water mixture slightly wets the chip. The contact angle with the glass wall is within 75\textdegree{} to 80\textdegree{} and 75\textdegree{} to 95\textdegree{} with the silicon wall (see the SI for more details on the estimations of the contact angle). Therefore, as long as there was liquid at the back channel, the Hele-Shaw cell would remain filled. The replenishing flow would then prevent the back channel from emptying. We found the right volumetric flow by slightly under-filling the chip such that a meniscus was visible at the back channel (figure \ref{fig:SketchSetUp}b). A volumetric flux of about 0.43 \si{\micro\liter\per\minute} kept the meniscus stationary. In experiments with dye, we used a slightly larger flow 0.5 \si{\micro\liter\per\minute} which resulted in the chip eventually getting fully filled. However, there  was no noticeable overflow at the observation region. Furthermore, the edge of the chip helped to keep the interface pinned. The pressure at the inlet of the chip was estimated to be about 19 \si{\pascal} higher than at the evaporating edge. This pressure is smaller than the pressure drop across the meniscus at the evaporating edge ($10^2$ to $10^3$ \si{\pascal}), in accordance with no overflow (see SI for details on the calculation of the pressure drop). Furthermore, the edge of the chip helped to keep the interface pinned.

We used Rhodamine 6G (SigmaAldrich, dye content 99\%) and 1 \si{\micro\meter} particles (FluoSpheres, carboxilate-modified, Life Technologies) for visualization and for $\mu$-PIV (30 fps), respectively. Further details on the $\mu$-PIV experiments can be found in appendix \ref{app:Experimental details micro-PIV}.

After each experiment the chip was cleaned by immersing it in a bath of acetone, isopropyl alcohol, and ethanol sequentially. In between each bath we used a flow of nitrogen for drying. 

The solutions were made at a 50 wt\% ratio between Milli-Q water (produced by a Reference A+ system (Merck Millipore) at 18.2 M\textOmega cm (at 25\si{\degreeCelsius})) and ethanol (Boom;  100\%(v/v), technical grade), prepared by measuring the mass of each component using a balance (Secura 224-1S, Sartorius). The solution was not degassed before experiments.

\subsection{Evolution of the instability from the interface into the cell}

\begin{figure}
	
	\centerline{\includegraphics[width=\linewidth]{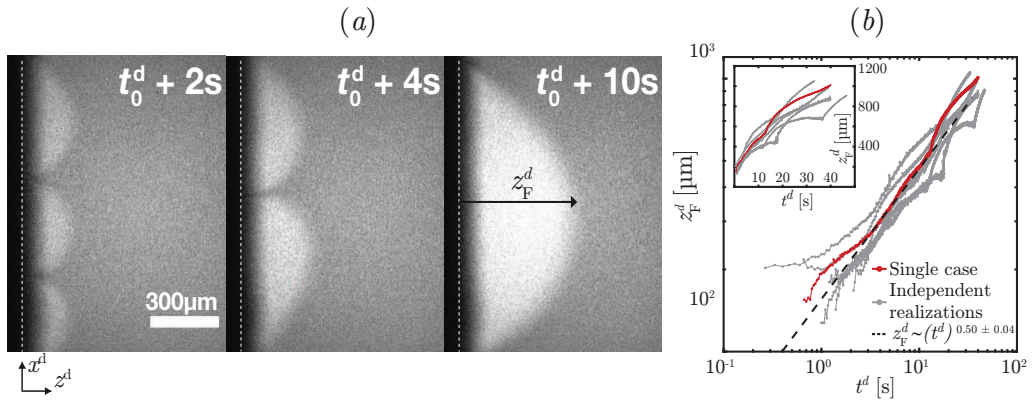}}
	\caption{(\textit{a}) Snapshots taken at the region marked with a red square in figure \ref{fig:SketchSetUp}(\textit{b}) at three different times. The snapshots show an evaporating 50wt\% ethanol-water solution dyed with Rhodamine 6G (See movie 1 of the SI for the whole time evolution). The dye accumulates in regions shaped as arches that grow and merge with each other over time. We have added the time $t_0^\mathrm{d}$ because the solution can start to evaporate even before it reaches the edge. The vertical white dashed lines indicate the position of the edge of the chip and the interface between air (left) and liquid (right). (\textit{b}) Height $z_\mathrm{F}^\mathrm{d}$ of the arches (see last panel of (\textit{a}) for definition) as function of time. Different independent realizations of the experiment are shown in grey, while a single case is highlighted in red. The black dashed line corresponds to the power law $z^\mathrm{d}_\mathrm{F} \propto (t^\mathrm{d})^{0.50\pm 0.04}$. All the realizations were fitted omitting the data for which $t^\mathrm{d}<1$ s. \textit{Inset:} Same plot, but with linear axes.}
	\label{fig:GrowInitialTimes}
\end{figure}

In one set of our experiments, we dyed the ethanol-water solutions with a trace amount of Rhodamine 6G (R6G). As evaporation takes place, non-volatile R6G accumulates close to the edge, resulting in brighter regions as can be seen in figure \ref{fig:GrowInitialTimes}(\textit{a}) (these snapshots were taken at the region marked with a red square in figure \ref{fig:SketchSetUp}(\textit{b})). These regions are shaped as arches of approximately same sizes, suggesting the presence of an instability \citep{Kollner_2015, Linde_1964_HS}. If this were not the case, we would expect the accumulation of R6G to be distributed uniformly all along the edge (white vertical dashed line), but not in specific regions.

Over time, the arches grow and merge with each other as seen in figure \ref{fig:GrowInitialTimes}(\textit{a}), where their number has gone from three to one over a few seconds. This kind of coarsening behaviour is expected for Marangoni instabilities \citep{Schwarzenberger_2014}. Such phenomenon was also observed in evaporation experiments inside a Hele-Shaw cell by \citet{Linde_1964_HS}. 

We measured the distance $z_\mathrm{F}^\text{d}(t)$ that the accumulated dye penetrated towards the inside of the cell as function of time (see third panel of figure \ref{fig:GrowInitialTimes}(\textit{a}) for an example of $z_\mathrm{F}^\text{d}$). Here the superscript $\text{d}$ indicates a dimensional quantity. To measure $z_\mathrm{F}^\text{d}$, we took the vertical average of the intensity field, resulting in an intensity profile as function of $z^\text{d}$. Then we took the derivative of this profile with respect to $z^\text{d}$. Finally, we looked for the distance from the edge where the derivative was zero within noise level and we assigned this value to $z_\mathrm{F}^\text{d}$. This process was then repeated for every time frame.

In figure \ref{fig:GrowInitialTimes}(\textit{b}), we show some examples of $z_\mathrm{F}^\text{d}(t)$ on a double logarithmic scale. A close inspection of the plots shows that during a merging event $z_\mathrm{F}^\text{d}$ accelerates and then slows down. However, as an overall trend, $z_\mathrm{F}^\text{d}$ grows effectively as a power law. Fourteen independent runs were fitted to $\propto (t^\mathrm{d})^{\alpha}$. An average exponent of $\alpha = 0.50 \pm 0.04$ suggests that $z_\mathrm{F}^\text{d}$ effectively grows in a diffusive-like manner. A similar behavior has been observed in liquid-liquid systems with Marangoni instabilities \citep{Kollner_2015}. Figure \ref{fig:GrowInitialTimes}(\textit{b}) includes cases with two different mass fractions of R6G: $10^{-4}$ and $10^{-5}$. There are no considerable changes in the exponent (see the SI for a plot with all cases).

We defined the time $t^\mathrm{d}=0$ \si{\second} to be the moment at which the solution reached the edge within our field of view. Although evaporation can take place while the chip gets filled, we expect its effect to be negligible as long as the liquid is far away from the edge. Nevertheless, it can trigger the instability already some time before the liquid reaches the edge. For the fits, data for which $t^\text{d}< 1$ s was omitted to minimize the effects of our uncertainty on the initial time. The rest of the time series was included in the fit.

\begin{figure}
	\centerline{\includegraphics{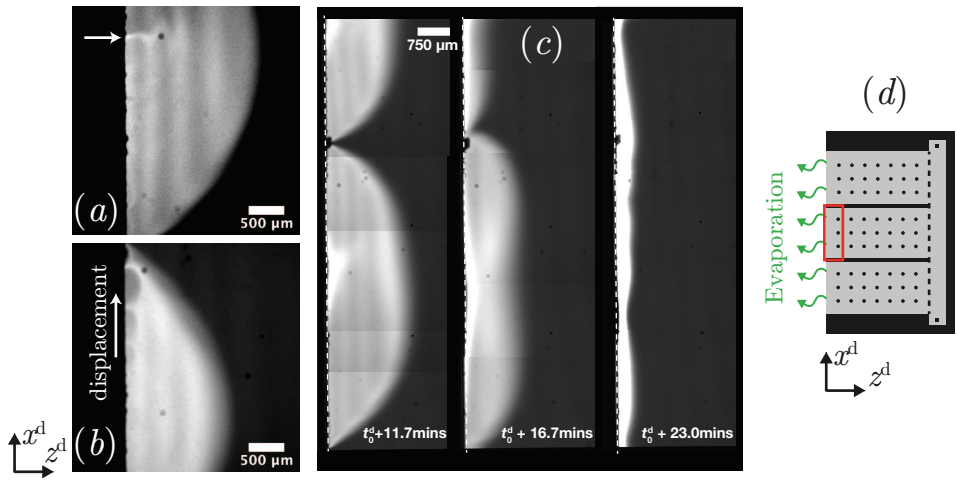}}
	\caption{Long-time flow dynamics for the same experiment as in figure 2. In (\textit{a}) we highlight how the dye flows back at the center of an arch in a meandering way (marked with a white arrow).  In (\textit{b}) secondary arches are clearly visible. They last for just a few seconds and travel towards the tip of the principal arch. For these two images we adjusted the contrast for visibility. In (\textit{c}) the arches have grown to a size comparable to the width of the  central part of the chip marked by the red square in (\textit{d}). The semi-arch visible in the first panel of (\textit{c}) is in contact with the upper wall. The white dashed line indicates the edge of the chip. The images in (\textit{c}) were made by stitching smaller pictures (taken over 20 s).}
	\label{fig:GrowLateTimes}
\end{figure}

As the arches get bigger, interesting phenomena take place inside them. We show in figure \ref{fig:GrowLateTimes}(\textit{a}) how the dye flowed back into the chip in a meandering way, and in figure \ref{fig:GrowLateTimes}(\textit{b}) short-lived secondary arches. The latter travelled towards the tips of the arches in a few seconds while changing in size, as also observed by \citet{Linde_1964_HS} and \citet{Schwarzenberger_2012}. In movie 2 of the SI the secondary arches are clearly visible for a case where the liquid was seeded with tracer particles.

Eventually, the lateral size of the arches becomes comparable with the width of the central part of the chip, as shown in the first panel of figure \ref{fig:GrowLateTimes}(\textit{c}). The area shown corresponds to the red square in the sketch of figure \ref{fig:GrowLateTimes}(\textit{d}). We did not observe arches spanning over this whole area. Instead the arches eventually start to reduce in size (second panel of figure \ref{fig:GrowLateTimes}(\textit{c})), sometimes causing the remaining ones to start traveling along the edge. Overall, after 10-25 min the instability fades away and the dye is swept to the interface (third panel of figure \ref{fig:GrowLateTimes}(\textit{c})). Then a narrow bright stripe parallel to the edge forms, suggesting that the gradient of surface tension is not strong enough to drive the flow anymore. 

A possible reason for the instability to stop is a reduced evaporation of ethanol. Indeed, after taking the chip out from the chamber, we observed that the instability can start again. Therefore, as the chip was moved to an environment without any ethanol vapour, evaporation of ethanol can be restarted and the instability triggered again.

\subsection{Visualization of the bulk flow by micro particle image velocimetry}

\begin{figure}
	\centerline{\includegraphics[width=\linewidth]{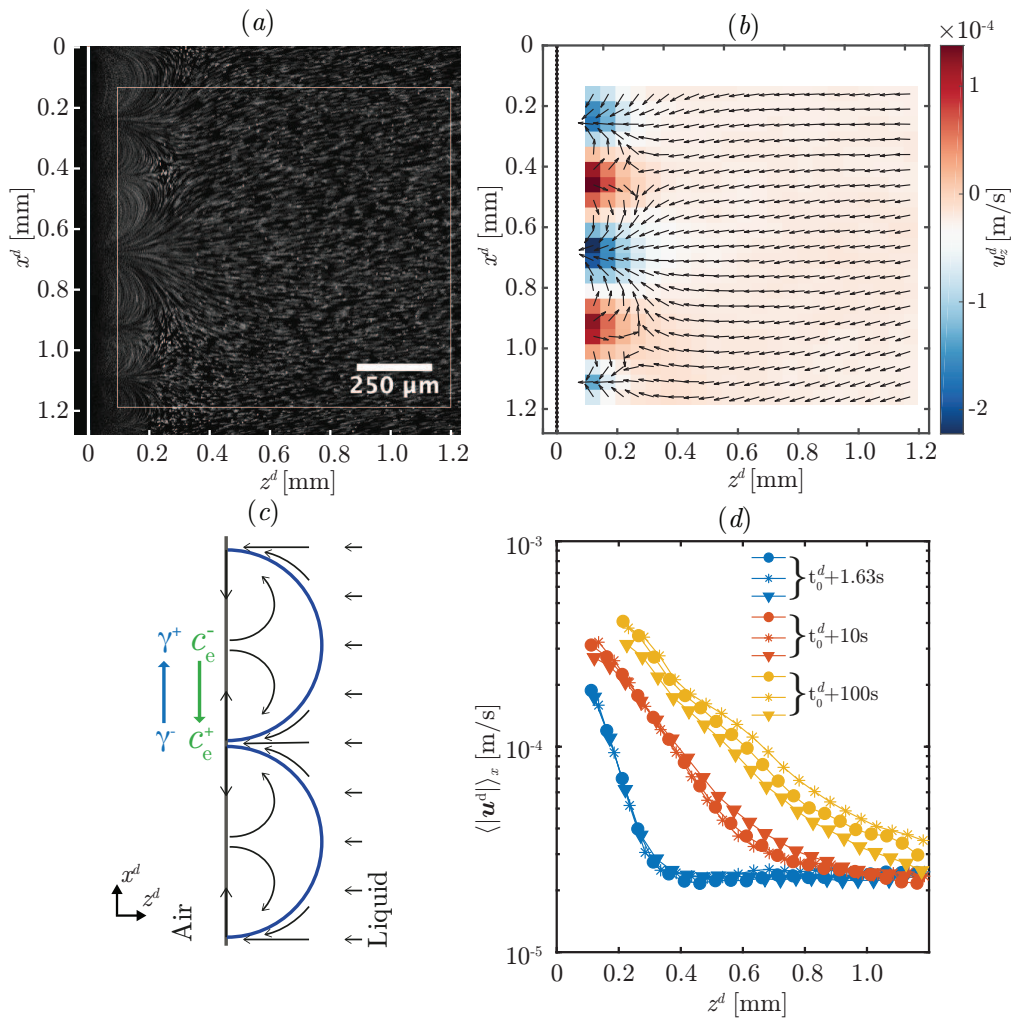}}
	\caption{Flow characterization while the instability is active. (\textit{a}) Streak lines formed by tracer particles obtained by averaging the intensity of 31 frames, allowing to see the same arches as with dye. As before the white dashed line indicates the edge. (\textit{b}) $\mu$-PIV measurement averaged over 1 s around $t^\mathrm{d} = 1.63$ s after the solution reached the edge. The measurement was taken inside the white square marked in (\textit{a}). The color code shows the horizontal component $u_z$ of the velocity to highlight the periodicity of the flow. The arrows were normalized to clearly show the direction (the vertical component and the total magnitude are shown in the SI. The black dotted line indicates the position of the edge. (\textit{c}) Sketch of the flow around and inside the arches. The gradient in concentration of ethanol produces a gradient in surface tension that drives a flow at the interface, resulting in convective rolls which cause the arch-like shape. (\textit{d}) Profiles of the velocity magnitude at different times in a semi-log scale. The profiles were obtained as a spatial average over the $x$ direction of the velocity fields. Different symbols represent different realizations to show reproducibility. One of the blue lines corresponds to the case at $t=1.63$ s shown in (\textit{b}). All cases where obtained from velocity fields averaged over 1 s around the time indicated by the legend.} 
	\label{fig:PIVSketch}
\end{figure}

To visualize the flow caused by the instability, the solution was seeded with 1 \si{\micro\meter} fluorescent particles at a very small volume fraction. In figure \ref{fig:PIVSketch}(\textit{a}) we show the streak lines made by the particles over a period of 1 s around 1.63 s after the solution reached the edge. Clearly shown in this figure, the arches are the result of advected dye (a typical video can be seen in movie 2 of the SI).

The direction of the flow can be read from panel (\textit{b}) of figure \ref{fig:PIVSketch} where we have plotted the corresponding normalized velocity vector field (averaged over 30 frames and a single experiment). The color code in this image corresponds to the horizontal component of the velocity, while the vertical component and the magnitude can be seen in the SI. We did not measure the velocity close to the edge for two reasons: One is the ``shadow'' caused by the edge of the glass plate. The other is the limited frame rate of the confocal system used for imaging, while closer to the edge the flow becomes too fast. 

We would expect the flow to be as depicted in figure \ref{fig:PIVSketch}(\textit{c}), where the velocity at the interface is directed towards the center of the arches. Then the fluid recirculates back to the tips of the arches. This kind of flows is typical for Marangoni driven systems \citep{Linde_1964_HS, Kollner_2015, Schwarzenberger_2014, Shi_2006}. Therefore, we can conclude that the flow changes direction at a distance smaller than the closest distance to the edge at which we can still see particles ($\sim 50$ \si{\micro\meter} from figure \ref{fig:PIVSketch}(\textit{a})).

The effect of the interfacial flow is quite dramatic as can be seen from figure \ref{fig:PIVSketch}(\textit{d}), where we show the mean velocity profiles (averaged over 30 frames and along the $x$ direction) for different times. From these plots we can see the huge difference in velocities close to the edge as compared to far away, going from some hundreds to about 20 \si{\micro\meter\per\second}. In particular, this decay of the velocity seems to occur exponentially as can be seen from the approximately linear character of the velocity profiles when plotted in a semi-logarithmic scale. With ongoing time the slopes get smaller and their extension larger, as result of the continuous growth of the arches. In fact the slope did not change as much at longer times, reflecting the square root dependence on time that we measured for $z_F^\mathrm{d}$. At longer times the arches are bigger than our field of view, so the profiles were obtained as an average over a fraction of an arch. Additionally, for the yellow curves we have masked the region inside the arch as the density of particles was too high for a correct measurement of the velocity (an example of a frame with the mask can be found in the SI). 

With the measured velocity we can now calculate a P{\'e}clet number, $\mathrm{Pe} = U^\mathrm{d}L^\mathrm{d}/D^\mathrm{d}_\mathrm{R6G}$, with $U^\mathrm{d}$ and $L^\mathrm{d}$ the typical velocity and length of the flow, and $D^\mathrm{d}_\mathrm{R6G}$ the diffusion coefficient of R6G in ethanol ($D^\mathrm{d}_\mathrm{R6G} = $ \si{\num{2.9e-10}\meter\squared\per\second} \citep{Hansen_1998}). For the typical velocity we can consider the maximum velocity inside the rolls (that we can measure) which is in the order of $10^{-4}-10^{-3}$ \si{\meter\per\second}. The typical size of the arches is in the order of $10^{-4}-10^{-3}$ \si{\meter}. This results in a range of $\mathrm{Pe} \sim 10 - 10^3$, clearly meaning that advection overcomes diffusion. Therefore, the dyed arches should result from the Marangoni rolls. As a last note, if we take the velocity far away from the arches ($\sim$20 \si{\micro\meter\per\second}) as the typical velocity caused by evaporation and the size of the arches just after the stability is over, namely about 2 mm, then we obtain $\mathrm{Pe}\sim 100$. The dye should indeed accumulate very close to the edge as advection overcomes the diffusion of the dye.


\section{Model equations and their numerical simulations}

\subsection{Model equations}\label{sec:ModelEqns}

\begin{figure}
	\centerline{\includegraphics[width=0.8\linewidth]{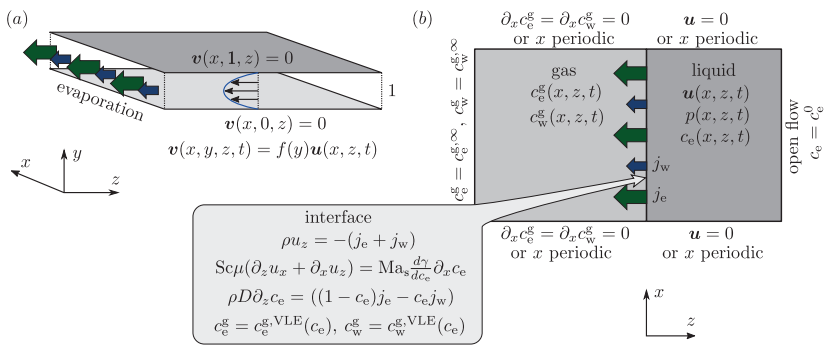}}
	\caption{Schematic of the numerical simulation. (\textit{a}) The small distance $\delta^\mathrm{d}$ between the plates of the Hele-Shaw cell allows for the assumption of a parabolic flow profile in $y$-direction, which effectively reduces the problem to two spatial dimensions $(x,z)$. (\textit{b}) Sketch of considered domains, fields and boundary conditions.}
	\label{fig:ModelSketch}
\end{figure}

In the following, the governing equations modelling the evaporation of a confined ethanol-water solution into humid air as used for numerical simulations are described. A schematic of the simulated system confined by two walls is shown in figure \ref{fig:ModelSketch}.

The equations are made non-dimensional by using the distance between plates $\delta^\mathrm{d}$ as a length scale, the initial mass density of the liquid $\rho_0^\mathrm{d}$ as a mass per volume scale, and the initial kinematic viscosity $\nu_0^\mathrm{d} = \mu_0^\mathrm{d}/\rho_0^\mathrm{d}$ to construct the viscous time scale $(\delta^\mathrm{d})^2/\nu_0^\mathrm{d}$. Therefore, the non-dimensional space coordinates, time, and velocities are given by $\boldsymbol{r} = \boldsymbol{r}^\mathrm{d}/\delta^\mathrm{d}$, $t = t^\mathrm{d}\nu_0^\mathrm{d}/(\delta^\mathrm{d})^2$, and $\boldsymbol{v} = \boldsymbol{v}^\mathrm{d}\delta^\mathrm{d}/\nu_0^\mathrm{d}$, respectively. Due to the presence of a binary solution in the liquid phase, the mass density $\rho = \rho^\mathrm{d}/\rho_0^\mathrm{d}$, the dynamic viscosity $\mu = \mu^\mathrm{d}/\mu_0^\mathrm{d}$, the diffusion coefficient $D = D^\mathrm{d}/\nu_0^\mathrm{d}$, and the surface tension $\SurfTension=\SurfTension^\mathrm{d}(\MassFractEth)/(\DerSurfTenMassFracDim)$ of the liquid are dependent on the local ethanol mass fraction $\MassFractEth(\boldsymbol{r},t)$. The surface tension is made non-dimensional using its derivative with respect to the mass fraction at $t=0$. It is the derivative and not the absolute value of the surface tension what determines the strength of the tangential stress. The non-dimensional mass transfer rate is given by $j_\alpha = j_\alpha^\mathrm{d}\delta^\mathrm{d}/(\nu_0^\mathrm{d}\rho_0^\mathrm{d})$ for $\alpha = \mathrm{e},\mathrm{w}$. As mentioned before, we use the superscript d in case we want to make reference to a dimensional variable. The subscript 0 indicates the value of a quantity at $t=0$.

This gives rise to the three-dimensional Navier-Stokes equations with varying density and viscosity:
\begin{eqnarray}
\partial_t \rho + \bnabla \cdot \left(\rho \boldsymbol{v}\right)&=&0, \label{eq:num:conti} \\
\rho\left(\partial_t \boldsymbol{v} + \boldsymbol{v}\bcdot\bnabla\boldsymbol{v}\right)&=&-\bnabla p +\bnabla\bcdot\left[\mu\left(\bnabla\boldsymbol{v}+\bnabla\boldsymbol{v}^\text{t}\right)\right]\,, \label{eq:num:nsthreedim} 
\end{eqnarray}
along with the convection-diffusion equation of the ethanol mass fraction $\MassFractEth$ with the composition-dependent diffusivity $D$:
\begin{eqnarray}
\rho \left(\partial_t \MassFractEth + \boldsymbol{v}\bcdot \bnabla\MassFractEth\right)=\bnabla\bcdot\left(\rho D\bnabla\MassFractEth\right)\, . \label{eq:num:advdiffthreed}
\end{eqnarray}

In the above equations we have neglected the effects of gravity, given the small distance between the plates $\delta^\mathrm{d}$ (20 \si{\micro\meter}), which is the relevant length for gravity in a horizontally oriented cell. In particular if we calculate the Galilei number $\mathrm{Ga} = g(\delta^\mathrm{d})^3(\rho_0^{\mathrm{d}}/\mu_0^{\mathrm{d}})^2$ and the Archimedes number $\mathrm{Ar}=g(\delta^\mathrm{d})^3\rho_0^\mathrm{d}\Delta\rho_0^\mathrm{d}/(\mu_0^{\mathrm{d}})^2$ based on the plate distance, the density difference between pure ethanol and water (207.7 \si{kg/m^3} at 293.15 \si{K} and 211.3 \si{kg/m^3} at 298.15 \si{K}), the dynamic viscosity (3.1 mPas at 293.15 \si{K} and 2.4 mPas at 298.15 \si{K}), and the density (0.918 \si{kg/m^3} at 293.15 \si{K} and 0.911 \si{kg/m^3} at 298.15 \si{K}) at 50 wt\%, we obtain $\mathrm{Ga} \sim 0.01$ and $\mathrm{Ar} = 0.0016-0.0026$. In both cases, their magnitude is much smaller than one, meaning that we can neglect gravity effects \citep{yu2015gravitational, Li2019}. 

As schematically depicted in figure \ref{fig:ModelSketch}(a), the three-dimensional flow inside the cell is strongly confined in the $y$-direction and enclosed by no-slip boundary conditions at the top and bottom plate. To the lowest non-trivial order compatible with these boundary conditions, a parabolic velocity distribution in $y$-direction can hence be assumed, i.e. $\boldsymbol{v}(x,y,z,t)=f(y)\boldsymbol{u}(x,z,t)$ with $f(y)=(by)(1-y)$ (see \ref{fig:ModelSketch}(a)). The two-dimensional velocity $\boldsymbol{u}$ is defined in the $(x,z)$-plane, i.e. $\boldsymbol{u}\bcdot\boldsymbol{e}_y=0$, and the constant $b$ is set to $b=6$ in the numerics so that the lateral velocity $\boldsymbol{u}$ coincides with the height average of the three-dimensional velocity $\boldsymbol{v}$. 
With these assumptions, the three-dimensional Navier-Stokes equations (\ref{eq:num:nsthreedim}) are simplified to two dimensions, resulting in a factor $6/5$ for the non-linearity and a drag term $-12\mu\boldsymbol{u}$ stemming from the shear of the parabolic flow profile \citep{Bizon1997_Chaos,Bratsun2004}:
\begin{eqnarray}
\partial_t \rho + \bnabla_\parallel \cdot \left(\rho \boldsymbol{u}\right)&=&0 \label{eq:num:contitwodim} \\
\rho\left(\partial_t \boldsymbol{u} + \frac{6}{5}\boldsymbol{u}\bcdot\bnabla_\parallel\boldsymbol{u}\right)&=&-\bnabla_\parallel p +\bnabla_\parallel\bcdot\left[\mu\left(\bnabla_\parallel\boldsymbol{u}+\bnabla_\parallel\boldsymbol{u}^\text{t}\right)\right]-12\mu\boldsymbol{u}\,. \label{eq:num:nstwodim} 
\end{eqnarray}
Here, $\bnabla_\parallel=(\partial_x,\partial_z)$ is the two-dimensional nabla operator. The liquid properties $\rho$, $\mu$ and $D$ are still allowed to depend on the local ethanol mass fraction, which is now also considered in its two-dimensional height-averaged projection, i.e. $\MassFractEth(x,z,t)$, which is subject to the two-dimensional convection-diffusion equation
\begin{equation}
\rho \left(\partial_t \MassFractEth + \boldsymbol{u}\bcdot \bnabla_\parallel\MassFractEth\right)=\bnabla_\parallel\bcdot\left(\rho D\bnabla_\parallel\MassFractEth\right)\, . \label{eq:num:advdifftwod}
\end{equation}
Considering only the two-dimensional projection of the concentration field $\MassFractEth$ does not take into account diffusion processes in $y$-direction, which can, in cooperation with the parabolic flow profile, contribute to an effectively enhanced projected two-dimensional diffusivity (cf. Taylor-Aris dispersion). However, due to the non-trivial and time-dependent flow, this effect cannot be taken into account within this model.

The gas phase is modeled as a mixture of air, ethanol, and water, for which we assume quasi-stationary vapour diffusion, i.e. we only consider the Laplace equations 
\begin{equation}
	\nabla^2 \VaporField_\text{e}=0 \qquad\text{and}\qquad\nabla^2 \VaporField_\text{w}=0
\end{equation}
for the vapour mass fractions $\VaporField_\text{e}$ and $\VaporField_\text{w}$ of ethanol and water, respectively. The superscript g indicates the gas phase. Then the mass fraction of air is given by $c_\mathrm{a}^\mathrm{g} = 1 - \VaporField_\text{e} - \VaporField_\text{w}$. The assumption of a quasi-stationary model has been validated previously for evaporating ethanol-water sessile droplets. In particular, the consideration of convection \citep{Diddens2017} or the full diffusion equation \citep{Diddens2017modeling} in the gas phase resulted in negligible differences in the evaporation rates or tangential velocities at the surface of the droplets. Given that the velocities in our experiments are smaller or in the same order as in the droplet case, we have decided to follow the same approach.

Opposed to the experimental set-up, the gas phase in the numerics is considered in a two-dimensional projection as well. To mimic the evaporation in a three-dimensional space, the length of the gas domain in $z$-direction is adjusted until the typical evaporation velocity matches the experimentally observed one. 

The boundary conditions for the velocity and for $\MassFractEth$ are shown in figure \ref{fig:ModelSketch}(b), namely an open stress free inflow at the liquid boundary far away from the liquid-gas interface and no-slip boundary conditions at the side walls. Alternatively, when only a section of the cell with some distance to the side walls is of interest, periodic boundary conditions in $x$-direction can be used (if so, this will be mentioned in the text). At the liquid-gas interface, ethanol and water evaporate with local non-dimensional mass transfer rates $j_\text{e}$ and $j_\text{w}$, respectively. This, together with the assumption of a static interface, yields the kinematic boundary condition
\begin{equation} \label{eqn:BC_Vel_fluxes}
\rho \VeloZ=-\left(j_\text{e}+j_\text{w}\right),
\end{equation}
where we have neglected the solubility of air in the binary solution, such that the air flux at the interface is zero. The negative sign arises because we have defined the fluxes positive when directed away from the liquid. 

Due to the small viscosity ratio of the gas phase with respect to the liquid phase, the varying interfacial tension induces a Marangoni shear stress only on the liquid phase, i.e.
\begin{equation}\label{eqn:BC_TangentialStress}
\mu\left(\partial_z \VeloX+\partial_x \VeloZ\right)= \frac{\MarangoniNumber_\mathrm{s}}{\SchmidtNumber} \frac{d\SurfTension}{d\MassFractEth}\partial_x \MassFractEth\,,
\end{equation}
where we have introduced the solutal Marangoni number  \citep{MACHRAFI2010} 
\begin{equation}\label{eqn:Ma}
	\MarangoniNumber_\mathrm{s} = -\frac{\delta^\mathrm{d}}{\mu_0^\mathrm{d}D_0^\mathrm{d}}\frac{d\SurfTension_0^\mathrm{d}}{d\MassFractEth},
\end{equation}
and the Schmidt number 
\begin{equation}\label{eqn:Ma}
	\mathrm{Sc} =\frac{\nu_0^\mathrm{d}}{D_0^\mathrm{d}},
\end{equation}
with $D_0^\mathrm{d}$ the diffusion coefficient of the liquid at time zero.

In the following section, for linear stability analysis we will introduce linear mass fraction profiles close to the edge, with slope $E$. Therefore, a characteristic change in mass fraction $\Delta c_\delta = E^\mathrm{d}\delta^\mathrm{d} = E$ over a distance equal to $\delta^\mathrm{d}$ is obtained. Then a ``modified'' Marangoni number \citep{MACHRAFI2010} 
\begin{equation}\label{eqn:MaStar}
\MarangoniNumber_\mathrm{s}
^*= \MarangoniNumber_\mathrm{s}\Delta c_\delta =  -\frac{(\delta^\mathrm{d})
^2 E^\mathrm{d}}{\mu_0^\mathrm{d}D_0^\mathrm{d}}\frac{d\SurfTension_0^\mathrm{d}}{d\MassFractEth}
\end{equation}
 can be introduced, which determines the stability of system. Notice that dispersion can result in an enhanced effective diffusion coefficient, and reduce the value of the Marangoni number. But as mentioned before, the determination of this effect is not trivial for our flow.

Due to the preferential evaporation of ethanol, the convection-diffusion equation (\ref{eq:num:advdifftwod}) is subject to the boundary condition 
\begin{equation} \label{eq:fluxBCComplete}
\rho D\partial_z \MassFractEth=((1-\MassFractEth)j_\text{e}-\MassFractEth j_\text{w})\,.
\end{equation}

The far-field boundary condition of the gas composition is set to that of ambient air, plus an optional relative humidity. At the liquid-gas interface, vapour-liquid equilibrium is imposed, i.e.
\begin{equation}
\VaporField_\text{e}=\VaporFieldVLE_\text{e}(\MassFractEth) \qquad\text{and}\qquad \VaporField_\text{w}=\VaporFieldVLE_\text{w}(\MassFractEth)\,,
\end{equation}
where the equilibrium is calculated by Raoult's law generalised by activity coefficients, i.e.
\begin{equation}\label{eqn:RaoultsLaw}
\VaporFieldVLE_\alpha(\MassFractEth)=\ActivityCoeff_\alpha(\MassFractEth)\MoleFract_\alpha(\MassFractEth)\frac{\SatPressure{\alpha}^\mathrm{d}(T)\MolarMass_\alpha^\mathrm{d}}{R^\mathrm{d}T^\mathrm{d}\rho^{g,d}}.
\end{equation}
Here, $\ActivityCoeff_\alpha$ are the activity coefficients, $\MoleFract_\alpha$ the liquid mole fractions, $\MolarMass_\alpha^\mathrm{d}$ the molar masses, and $\SatPressure{\alpha}^\mathrm{d}$ the saturation pressures of the pure components, whereas $R^\mathrm{d}$ is the universal gas constant, $T^\mathrm{d}$ is the temperature of the system, and $\rho^{g,d}$ is the density of the gas. Notice that the equilibrium mass fraction changes in time through $\MoleFract_\alpha(\MassFractEth(z=0,x,t))$ and  $\ActivityCoeff_\alpha(\MassFractEth(z=0,x,t))$.
The evaporation rates are finally given by the diffusive vapour flux at the interface, i.e.
\begin{equation}\label{eqn:EvaporationRates}
j_\text{e}=\GasDensity\VaporDiffusivity_\text{e}\partial_z \VaporField_\text{e}/\SchmidtNumber \qquad\text{and}\qquad j_\text{w}=\GasDensity\VaporDiffusivity_\text{w}\partial_z \VaporField_\text{w}/\SchmidtNumber\,,
\end{equation}

where $\GasDensity = \rho^{g,d}/\rho_0^\mathrm{d}$ and $\VaporDiffusivity_\alpha = D^\mathrm{g,d}_\alpha/D_0^\mathrm{d}$ are the gas to liquid density and diffusivity ratios.

As we mentioned in our introduction, we do not consider thermal effects, which in the experiments can take place by means of evaporative cooling. In order to justify this assumption we refer to \citet{Linde_1964_HS} and \citet{MACHRAFI2010}, who found that the solutal effect is dominant. In particular, \citet{MACHRAFI2010} showed that the ratio between solutal and thermal Marangoni numbers is large for their ethanol-water solutions. We can obtain the same ratio in our case considering the thermal Marangoni number as
\begin{equation}
	\mathrm{Ma}_\mathrm{th} = \frac{\delta^\mathrm{d} \Delta T^\mathrm{d}} {\mu_0^\mathrm{d}D_{0,T}^\mathrm{d}} \frac{d\SurfTension_0}{dT^\mathrm{d}},
\end{equation}
where $T^\mathrm{d}$ is the temperature, $\Delta T^\mathrm{d}$ is a characteristic temperature change along the interface and $D_{0,T}^\mathrm{d}$ is the thermal diffusion coefficient. The specific values used are shown in table \ref{tab:DimensionalValues}.

We do not have the values of the characteristic temperature change, but when assuming the range is between 1 K and even as unrealistically high as 10 K, the ratio $\MarangoniNumber_\mathrm{s}^*/\mathrm{Ma}_\mathrm{th} \sim 10^3 - 10^4$. Therefore, the solutal Marangoni effect is much stronger than the thermal one. This is consistent with the case of evaporating sessile ethanol-water droplets, where thermal Marangoni effects could be neglected as long as the ethanol has not yet fully evaporated \citep{Diddens2017}.

\subsection{Numerical simulation implementation}\label{sec:NS_implementation}

For the numerical simulation we used a finite element method implemented on the basis of the finite element package \textsc{oomph-lib} \citep{Heil2006}. 
The liquid in the Hele-Shaw cell and the adjacent gas domain are discretised by a rectangular mesh, using triangular Taylor-Hood elements for the degrees of freedom of the velocity and pressure space and linear shape function for the composition in both domains. The composition-dependence of the liquid and gas properties are obtained by models or fits of experimental data (see \citet{Diddens2017} for details). The general numerical technique is described in more detail and compared with various experiments in e.g. \citet{Diddens2017,Diddens2017c,Li2019,Li_2019_Bouncing}. In the present case, however, the method is generalised by considering the drag term in the momentum equation (\ref{eq:num:nstwodim}).

In all simulations the number of triangular elements was the highest close to the interface and then reduced continuously the further away from the interface. More details on the meshes used for different simulations can be find in the SI.

For the simulation used for comparison with experiments (section \ref{sec:NumResultsLarge}) the domain size was selected to be the same as the observation area in experiments in the vertical direction. This was already quite demanding computationally, so we used a reduced resolution such that the simulation time would not be exceedingly long. However, this came at the price that the boundary condition (\ref{eq:fluxBCComplete}) was not fully satisfied due to the steep gradient at the edge which could not be captured by our linear elements. Nonetheless, we were able to observe a good agreement with experiments. More details can be seen in the SI. 
In the case of simulations used for comparison with linear stability analysis, we made use of three different configurations of the mesh which we describe in the SI. 

In all simulations the initial mass fraction was $\MassFractEth = 0.5$, the relative humidity far away was taken equal to 50\% and the temperature was set to 295.15 \si{K}. The instability was triggered by numerical noise in all cases. To process the data, we linearly interpolated the mass fraction and velocity fields from the unstructured mesh used for simulations into a regular rectangular mesh.

\subsection{Numerical results and comparison with experiments} \label{sec:NumResultsLarge}

\begin{figure}
	
	\centerline{\includegraphics[width=\linewidth]{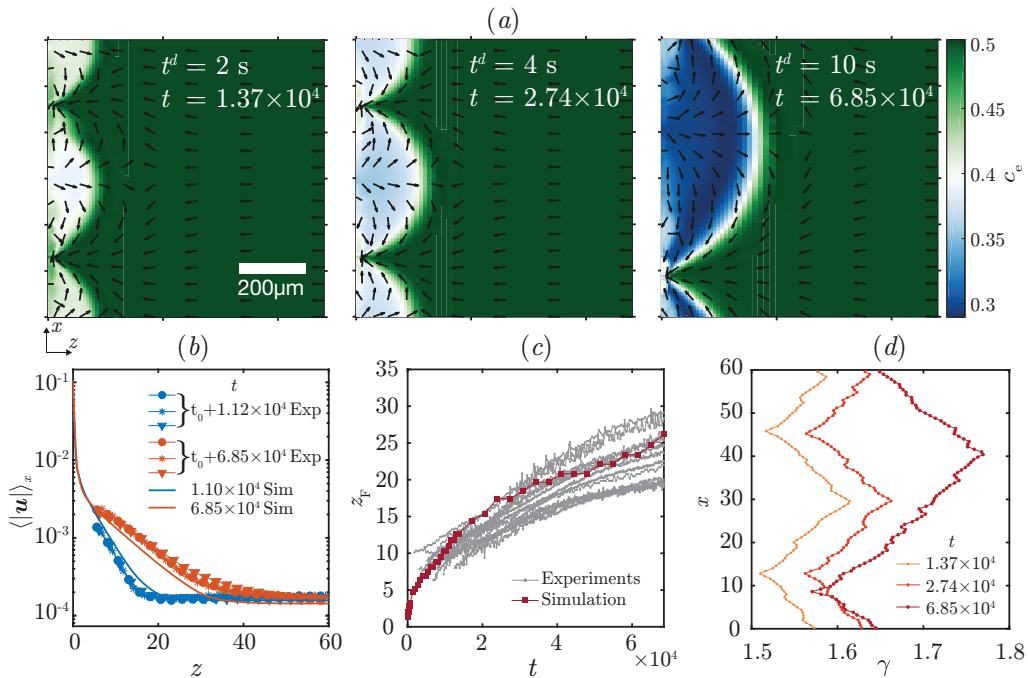}}
	\caption{ (\textit{a}) Ethanol mass fraction field on the liquid phase $\MassFractEth$  obtained from a simulation with periodic boundary conditions. The times are the same as in figure \ref{fig:GrowInitialTimes}(\textit{a}). The normalized velocity vector field is plotted on top of the concentration field. As in experiments, with progressing time, the arches merge with each other. Close to the edge, smaller secondary arches become visible (a video showing the whole time evolution can be seen in movie 3 of the SI). (\textit{b}) Averaged velocity profiles at two different times. The curves with markers show the same profiles as in figure \ref{fig:PIVSketch}(\textit{d}) where different markers indicate different runs. The solid lines are results from the simulation. For the blue curves, we took the closest time obtained from simulation. (\textit{c}) Height $z_F$ of the arches as function of time. The gray dotted lines show all the experimental realizations up to $t = $ \si{\num{6.85e4}} ($t^\mathrm{d} = 10$ \si{\second}). The dark red line with squares shows the results from the simulation. (\textit{c}) Surface tension $\gamma$ along the edge for the three times shown in (\textit{a}). The small scale jumps in the curves originate from the smaller secondary rolls.}
	\label{fig:SimAll}
\end{figure}

In figure \ref{fig:SimAll} we show results obtained from a simulation using periodic boundary conditions and a simulation with a domain size equal to the field of view that we had for experiments. Other conditions can be found in table \ref{tab:DimensionalValues} in Appendix \ref{app:TableOfValues} and in the SI. The three snapshots of figure \ref{fig:SimAll}(\textit{a}) show the evolution of the instability over three different times. As opposed to experiments, this time we can directly observe the ethanol concentration field and on top the normalized velocity vector field. It is clear that Marangoni rolls advect ethanol-depleted fluid into the arch shaped regions. As in experiments, the arches merge and grow over time.

After the arches are large enough we can see secondary arches too. From movie 3 in the SI it can be seen that initially only the principal arches are present. Inside the arches, ``pockets'' of rich ethanol solution are present. Secondary rolls can appear from these pockets. We can think of this as if locally the concentration field is back to the initial state when no instability was present and the gradient of concentration just builts up. Therefore, besides the fact that now there is a flow along the interface, locally a second instability can be triggered as described by \citet{Kollner_2013_multiscale} for 3D simulations of a liquid-liquid interface.

The length of the gas domain in the $z$ direction was selected such that the velocity far away from the interface would be close to the velocity measured in experiments. We show this matching in figure \ref{fig:SimAll}(\textit{b}). We plot non-dimensional averaged velocity profiles obtained at two different times, both from experiments and from the simulation. The curves with markers are experimental results, corresponding to the two earliest cases that were already shown in figure \ref{fig:PIVSketch}(\textit{d}). The solid lines represent the results obtained from a single simulation, a single time and averaging along the $x$ direction. (for the blue curve, we took a profile at the closest time obtained from the simulation).  Notice that the complete profiles agreed quite well with the experimental measurements, despite the fact that we only matched the velocity far away from the edge.

It is important to note that the merging between arches did not take place at the same times between simulations and experiments, causing a discrepancy between the experimental and numerical velocity profiles during certain periods of time. In figures 10 and 11 of the SI, we make more detailed comparisons of the velocity between experiments and simulations. We have noticed that the velocity increases sharply during a merging event, but changes slowly otherwise. This is in agreement with the faster growth of $z_\mathrm{F}$ during merging. Moreover, as long as single arches have similar sizes, the local velocity is approximately the same. In figure 11 of the SI, we compare the experimental velocity profiles of arches of similar size at different times or from different realizations. The velocity profiles match with each other. The velocity profiles obtained from a simulated arch with a similar size compare well, too. On the other hand, the local velocities did not match between experiments and simulations if the arches had different sizes. 

For $t = 6.85 \times 10^4$, the velocity far away from the interface is smaller than in experiments. This could be a result of the comparable size of the arch with the simulation domain. Nevertheless, the main features stay similar as in experiments with a region of exponential decay until the velocity reaches the far end value. Furthermore, we can now see that the velocity close to the edge (i.e., small values $z\lesssim 2$) can reach much larger values than those measured experimentally, about two orders of magnitude larger.

We tracked the height $z_\mathrm{F}$ as in experiments by measuring the distance from the edge at which the vertically averaged profile of the mass fraction had reached the far field mass fraction. The result of this process is shown in figure \ref{fig:SimAll}(\textit{c}). Both experimental and simulation results are plotted together, again showing a good match despite our uncertainty in the initial time in the experiments. Note that there is no adjustable parameter, we only matched the initial far field velocities in the liquid.

From the simulations we can also have access to the local surface tension $\gamma(x)$ as can be seen in figure \ref{fig:SimAll}(\textit{d}). We show the surface tension $\gamma(x)$ corresponding to the three snapshots in figure \ref{fig:SimAll}(\textit{a}). The curves confirm the presence of gradients going from the center of the arches to their meeting points. Overall, the gradient in surface tension seems to stay approximately constant over the time span shown here. The small scale jumps in the curves originate from the secondary rolls.

Before a merging event, both in simulations and in experiments, the stagnation point at the summit of the arches displaces towards one side. Therefore,  the flow towards the edge in between two arches  reduces. In turn, the supply of ethanol-rich solution also decreases. A similar situation was observed by \citet{Kollner_2013_multiscale}, where the bulk flow caused by the Marangoni rolls hindered the flow towards the interface. Afterwards, the two adjacent rolls merged into one.

\section{Linear stability analysis}\label{MainSec:LSA}

\subsection{Simplified model}

In this section we will perform a linear stability analysis of a simplified version of the model equations described in section \ref{sec:ModelEqns}. In this way we will be able to obtain analytical solutions that are very close to those obtained by \citet{SternlingScriven1960AIChE}, with the addition of the drag term which originates from the wall friction.

In this simplified version of our model equations we will take the density, dynamical viscosity and diffusion coefficients as constants with respect to mass fraction as has been done for different systems before with good results \citep{Bratsun2004,MACHRAFI2010, Bizon1997_Chaos}. Furthermore, we will assume that surface tension is a linear function of the mass fraction such that the derivative of the surface tension with respect to the mass fraction is constant.
In particular we will take the values at time zero such that $\rho^\mathrm{d}=\rho_0^\mathrm{d}$, $\mu^\mathrm{d} = \mu_0^\mathrm{d}$, $D^\mathrm{d} = D_0^\mathrm{d}$, and $d\SurfTension^\mathrm{d}/d\MassFractEth=\DerSurfTenMassFracDim$. In other words $\rho=\mu=\DerSurfTenMassFracNonDim=1$, and $D = 1/\SchmidtNumber$,  which results in the simplified model equations for the liquid phase
\begin{eqnarray}
\bnabla_\parallel \cdot  \boldsymbol{u}&=&0 \label{eq:LS:contitwodim} \\
\partial_t \boldsymbol{u} + \frac{6}{5}\boldsymbol{u}\bcdot\bnabla_\parallel\boldsymbol{u}&=&-\bnabla_\parallel p +\bnabla_\parallel^2\boldsymbol{u}-12\boldsymbol{u}\, \label{eq:LS:nstwodim} \\
\partial_t \MassFractEth + \boldsymbol{u}\bcdot \bnabla_\parallel\MassFractEth&=&\frac{1}{\SchmidtNumber}\bnabla_\parallel^2\MassFractEth\, \label{eq:LS:advdifftwod}
\end{eqnarray}
and for the gas phase
\begin{equation} \label{eq:LS:Laplacetwod}
\nabla^2 \VaporField_\text{e}=0 \qquad\text{and}\qquad\nabla^2 \VaporField_\text{w}=0.
\end{equation}

Similarly, the boundary condition for the tangential stresses at the interface (equation (\ref{eqn:BC_TangentialStress})) is simplified to
\begin{equation} \label{eq:SurfTenBC_LS}
\partial_z \VeloX+\partial_x \VeloZ= \frac{\MarangoniNumber_\mathrm{s}}{\SchmidtNumber} \partial_x \MassFractEth.
\end{equation}

Additionally, we neglect the velocity caused by evaporation such that $u_z(x,0,t)=0$. This assumption has been validated by comparing simulations with and without making the normal velocity equal to zero at the edge. The results for both cases were virtually the same. Therefore, for the linear stability analysis we kept $u_z(x,0,t)=0$, while for the simulations $u_z(x,0,t)\neq0$. In the experimental section we obtained P\'{e}clet numbers larger than 1 in odds with this assumption, however for the onset of instability a more proper length scale to be consider is $\delta^\mathrm{d}$ (the gap of the Hele-Shaw cell), which gives $\mathrm{Pe}\sim 1$ meaning that we are in the limit where we could actually neglect this velocity. 

After neglecting the normal velocity, from equation (\ref{eqn:BC_Vel_fluxes}) we have $j_\mathrm{e}+j_\mathrm{w} = 0$ at the interface, then equation (\ref{eq:fluxBCComplete}) simplifies to $j_\mathrm{e} = \partial_z \MassFractEth/\SchmidtNumber$. Therefore, introducing this expression for the mass transfer rate of ethanol into equation (\ref{eqn:EvaporationRates}), we get 
\begin{equation} \label{eq:fluxBCSimplified}
\partial_z\MassFractEth = \GasDensity\VaporDiffusivity_\text{e}\partial_z \VaporField_\text{e},
\end{equation}
recalling that $\VaporDiffusivity_\mathrm{e} = D^\mathrm{g,d}_\mathrm{e}/D_0^\mathrm{d}$.

Raoult's law is simplified by taking a Taylor expansion of the molar fraction around $\MassFractEth(t=0)\equiv c_\mathrm{e,0} \ne0$, such that $\MoleFract_\mathrm{e} = \left(1+(1/\MassFractEth-1)M_\mathrm{e}^\mathrm{d}/M_\mathrm{w}^\mathrm{d} \right)^{-1} \approx a_0(c_\mathrm{e,0},M_\mathrm{e}^\mathrm{d}/M_\mathrm{w}^\mathrm{d}) + a_1(c_\mathrm{e,0},M_\mathrm{e}^\mathrm{d}/M_\mathrm{w}^\mathrm{d}) \MassFractEth + O(\MassFractEth^2)$. Finally, if we take the activity coefficient $\ActivityCoeff_\alpha = 1$ we are left with
\begin{equation} \label{eq:Raoults_Expanded}
\VaporFieldVLE_e(\MassFractEth) \approx c_\mathrm{sat}(a_0 + a_1 \MassFractEth),
\end{equation}
where $c_\mathrm{sat} =  p_\mathrm{e,sat}^\mathrm{d}(T)M_\mathrm{e}^\mathrm{d}/(R^\mathrm{d}T^\mathrm{d}\rho^{g,d}_0)$. An analogous expression can be find for water.

\subsection{Linear stability of 2D equations}

For the base case we will consider no flow in the liquid phase, i.e. $\pmb{u}_0 = 0$, which results in a constant pressure $p_0$. While for the mass fraction linear profiles close to the edge will be taken, in the same spirit as \citet{SternlingScriven1960AIChE},
\begin{equation}\label{eqn:LinProfLiq}
c_\mathrm{e,0} = Ez+\mathcal{C}_0,
\end{equation}
\begin{equation}\label{eqn:LinProfGas}
c_\mathrm{e,0}^\mathrm{g} = E^\mathrm{g}z+\mathcal{C}^\mathrm{g}_0,
\end{equation}
with $E$ and $E^\mathrm{g}$ made non-dimensional with $\delta^\mathrm{d}$. A more complete model would consider a time evolving base state with an error-function-like shape, however, then the equation has to be solved numerically. Here we consider that close to the edge the profiles can be approximated as linear. We also assume that the changes in concentrations at the edge are small around the time the instability takes place. By focusing our analysis to the region close to the interface, our analysis corresponds to the small wavelength regime.

We considered small perturbations in the standard way for the velocity, pressure, and mass fraction fields, 
\begin{eqnarray}
\pmb{u}&=& \epsilon \pmb{\hat{u}}_1(z) e^{i\kappa x+\sigma t} \label{eqn:Pert_u}\\
p & = & p_0 + \epsilon \hat{p}_1(z) e^{i\kappa x+\sigma t} \label{eqn:Pert_p}\\
\MassFractEth & = & c_\mathrm{e,0}(z,t) + \epsilon \hat{c}_\mathrm{e,1}(z) e^{i\kappa x+\sigma t} \label{eqn:Pert_cl}\\
\VaporField_\mathrm{e} & = & \VaporField_\mathrm{e,0}(z,t) + \epsilon \hat{c}^\mathrm{g}_\mathrm{e,1}(z) e^{i\kappa x+\sigma t} \label{eqn:Pert_cg},
\end{eqnarray}
where $\epsilon$ is a small number, $\kappa$ is the non-dimensional wavenumber of the perturbation, $\sigma$ its non-dimensional growth rate,  and $\pmb{\hat{u}}_1(z)$, $\hat{p}_1(z)$, $\hat{c}_\mathrm{e,1}(z)$, and $\hat{c}_\mathrm{e,1}^\mathrm{g}(z)$, the amplitudes of the corresponding perturbations in velocity, pressure, and mass fraction in the liquid and gas, respectively. These perturbed quantities can be used to linearize equations (\ref{eq:LS:contitwodim}) to (\ref{eq:LS:Laplacetwod}), resulting in the following dispersion relation (see appendix \ref{app:DerivationDispRelation} for further details):
\begin{eqnarray}\label{eqn:NonDimDispRel}
\MarangoniNumber_\mathrm{s}^* & = & \kappa^2 \left(c_\mathrm{sat}a_1\rho^\mathrm{g}\VaporDiffusivity_\text{e}+ \sqrt{1+\frac{\sigma}{\kappa^2}\SchmidtNumber}\right)\left(1+\sqrt{1+\frac{12}{\kappa^2} + \frac{\sigma}{\kappa^2}}\right)\nonumber\\
&&\left(1+\sqrt{1+\frac{\sigma}{\kappa^2}\SchmidtNumber}\right)\left(\sqrt{1+\frac{12}{\kappa^2} + \frac{\sigma}{\kappa^2}}  + \sqrt{1+\frac{\sigma}{\kappa^2}\SchmidtNumber}\right).
\end{eqnarray}

Notice that this is a simplified version of equation (30) from \citet{SternlingScriven1960AIChE}, but with the addition of the drag term and a Marangoni number $\MarangoniNumber_\mathrm{s}^*$ based on the plate distance $\delta^\mathrm{d}$.

From this equation we can obtain the condition of marginal stability by setting $\sigma=0$, resulting in 
\begin{equation}\label{eqn:MarginalStability}
	\kappa_{\sigma=0} =  \sqrt{\frac{\MarangoniNumber_\mathrm{s}^*}{8\mathcal{M}}} -3\sqrt{\frac{8\mathcal{M}}{\MarangoniNumber_\mathrm{s}^*}},
\end{equation}
where $\mathcal{M} = 1 + c_\mathrm{sat}a_1\rho^\mathrm{g}\VaporDiffusivity_\text{e}$. This relation is independent of $\mathrm{Sc}$, which was expected from the results of \citet{SternlingScriven1960AIChE}. Their results can be recovered in the limit of large $\MarangoniNumber_\mathrm{s}^*$, i.e. $\kappa_{\sigma=0} = (\MarangoniNumber_\mathrm{s}^*/8\mathcal{M})^{0.5}$ where the dependence on $\delta^\mathrm{d}$ is not present anymore, as $\kappa_{\sigma=0}\propto \delta^\mathrm{d}$ as well as $\sqrt{\MarangoniNumber_\mathrm{s}^*}\propto \delta^\mathrm{d}$. This is equivalent to making the gap between plates large enough so that the limit of an unbounded system is reached \citep{martin2002gravitational}.

We can now find a critical Marangoni number $\mathrm{Ma}^*_c\equiv24\mathcal{M}$ for which $\kappa_{\sigma=0} = 0$. Below this critical value, the right hand side of equation (\ref{eqn:MarginalStability}) becomes negative, while $\kappa$ has to be positive. Therefore for $0 < \MarangoniNumber_\mathrm{s}^* \leq \mathrm{Ma}^*_c$ the system should become stable for all wavelengths.  Furthermore, the Marangoni number can become negative if either the gradient $E$ or the derivative of surface tension with concentration $\DerSurfTenMassFracDim$ change in sign. Of course this is not possible as the wavenumber has to be a real positive quantity. However, this does not exclude the possibility of an oscillatory instability, which could be expected for negative Marangoni numbers considering the results from \citet{SternlingScriven1960AIChE}. Given that the evaporating ethanol-water system results in positive Marangoni numbers, we will restrict ourselves to this regime.

\begin{figure}
	\centerline{\includegraphics[width=\linewidth]{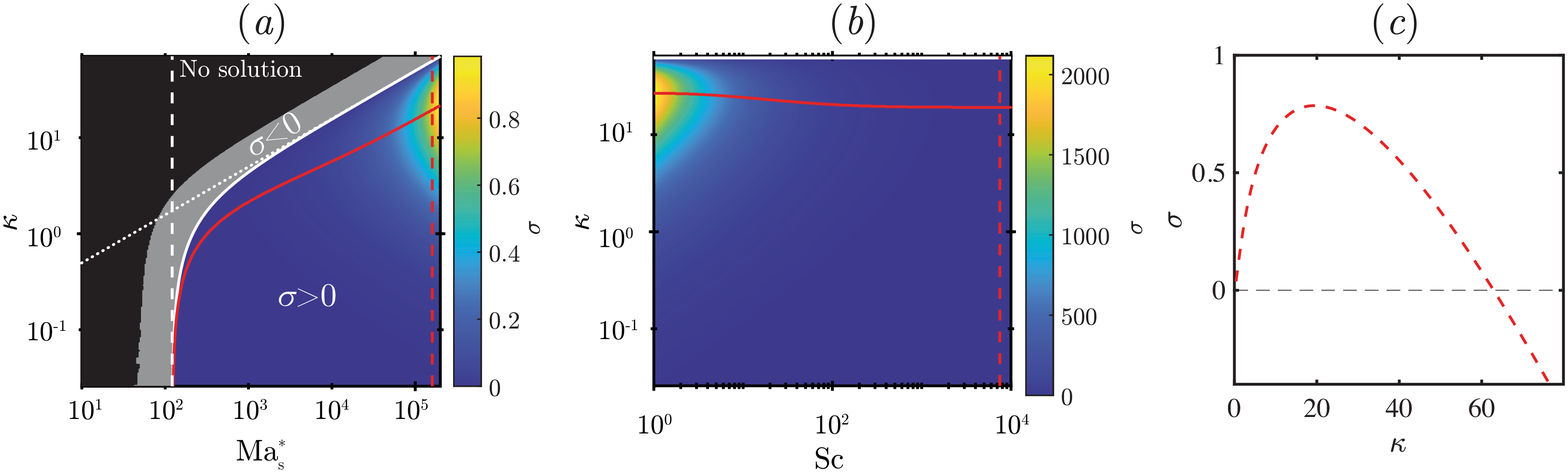}}
	\caption{Phase diagram of the growth rate $\sigma$ for different values of $\kappa$ and (\textit{a}) $\MarangoniNumber_\mathrm{s}^*$, with constant $\mathrm{Sc}=$ \si{\num{7.51e3}}, and different values of (\textit{b}) $\mathrm{Sc}$, with constant $\MarangoniNumber^*_\mathrm{s}=$ \si{\num{1.54e5}}. The vertical white dashed line in (\textit{a}) marks $\mathrm{Ma}^*_\mathrm{c}=121.2$, below which the system is always stable, while the dotted line corresponds to the limit $\kappa_{\sigma=0} = (\MarangoniNumber_\mathrm{s}^*/ 8\mathcal{M})^{0.5}$. In both (\textit{a}) and (\textit{b}), we have clamped negative values of $\sigma$ to zero to facilitate visualization (grey region). The solid white curve indicates $\sigma=0$, while the solid red line indicates $\mathrm{max}(\sigma)$. The black region in (\textit{a}) indicates where we did not find solutions for the dispersion relation. (\textit{c}) The growth rate $\sigma$ as function of $\kappa$ corresponding to the red vertical dashed line shown in (\textit{a}).}
	\label{fig:LinearInst_PhaseDiagram}
\end{figure}

\subsection{Numerical solution of the dispersion relation}
Equation (\ref{eqn:NonDimDispRel}) was solved numerically to obtain $\sigma$ as function of $\kappa$ and $\mathrm{Ma}^*_\mathrm{s}$ for $\SchmidtNumber = 7.51\times 10^3$, and the properties shown in table \ref{tab:DimensionalValues} in appendix \ref{app:TableOfValues} corresponding to an ethanol-water solution at 50 wt\% and 295 \si{\K}. The results can be seen in figure \ref{fig:LinearInst_PhaseDiagram}(\textit{a}), where we have clamped negative $\sigma$ to zero to facilitate visualization. Equation \ref{eqn:MarginalStability} is plotted as a solid white line, while the purely 2D limit, $\kappa_{\sigma=0} = (\MarangoniNumber_\mathrm{s}^*/ 8\mathcal{M})^{0.5}$ is highlighted by a dotted line. The red solid line corresponds to the locations where $\sigma$ is maximum for a given $\mathrm{Ma}^*$, i.e. the fastest growing modes. 

Indeed, for $\MarangoniNumber_\mathrm{s}^*<\mathrm{Ma}^*_\mathrm{c}=121.2$ the system is expected to be stable, as $\sigma<0$ for every value of $\kappa$. However, we must notice that for regions where $\sigma<-\kappa^2/\SchmidtNumber$, we did not find solutions (marked as a black region). This region corresponds to where three of the radicals become purely imaginary. We looked for imaginary solutions of $\sigma$, without success. While our search was not exhaustive, we are in the region where \citet{SternlingScriven1960AIChE} only found real solutions for the purely 2D case so we think it is reasonable not to find complex solutions.

Once a particular liquid-gas system has been fixed (with a given mass fraction gradient $E$), the Marangoni number $\MarangoniNumber_\mathrm{s}^*$ indicates the effect of the gap between the walls of the Hele-Shaw cell. By comparing the solid and dotted lines in figure \ref{fig:LinearInst_PhaseDiagram} we can observe that the effect of the confinement as compared to the purely 2D case is only visible for $\MarangoniNumber_\mathrm{s}^*\lesssim 1000$. Meaning that for larger values the purely two dimensional model of \citet{SternlingScriven1960AIChE} should be enough to calculate the critical wavelength. In other words, for a given set of properties, as long as $\delta^\mathrm{d}$ is large enough, but not so large that gravity starts to play a role, the use of the purely 2D model is sufficient.

According to \citet{SternlingScriven1960AIChE}, if the conditions $D^\mathrm{d}/D^\mathrm{g,d}<1$, $\nu^\mathrm{d}/\nu^\mathrm{g,d}<1$, and $d\SurfTension^\mathrm{d}/d\MassFractEth<0$ are satisfied, while solute transfer occurs from the liquid to the gas, the system should be unstable. However, consideration of the drag of the walls for an evaporating binary liquid introduces the possibility of stability as long as $\MarangoniNumber_\mathrm{s}^* < \MarangoniNumber^*_\mathrm{c}$.

We also looked at the effect of the Schmidt number, $\mathrm{Sc}$, as shown  in figure \ref{fig:LinearInst_PhaseDiagram}(\textit{b}) for a fixed $\MarangoniNumber_\mathrm{s}^*=$ \si{\num{1.54e5}}. As mentioned by \citet{SternlingScriven1960AIChE}, varying $\mathrm{Sc}$ shows only a minor effect which was already suggested by the fact that $\kappa_{\sigma=0}$ is not a function of this parameter. Only for $\mathrm{Sc}\lesssim50$ there is a slight increase on the value of the fastest growing wavenumber.

A plot of $\sigma$ as function of $\kappa$ is displayed in figure \ref{fig:LinearInst_PhaseDiagram}(\textit{c}). This case corresponds to the dashed red line in both panels (\textit{a}) and (\textit{b}) of figure \ref{fig:LinearInst_PhaseDiagram} ($\MarangoniNumber_\mathrm{s}^* =$ \si{\num{1.54e5}}  and $\SchmidtNumber =$ \si{\num{7.51e3}}). We have selected this particular case, because it corresponds to the simulated case in section \ref{sec:NumResultsLarge}. In order to know the right value of $E$, we conducted simulations in a smaller domain such that the mismatch of the boundary condition was not present. If we then assume that this corresponds to the same conditions as in experiments (for which we do not know the actual gradient), it would be in accordance with the presence of an instability, as the Marangoni number is above the critical value. 

With respect to the term $c_\mathrm{sat}a_1\rho^\mathrm{g}\VaporDiffusivity_\text{e}$, it affects the range over which the systems becomes stable, as clearly seen from equation (\ref{eqn:MarginalStability}). In particular, it increases the value of $\mathrm{Ma}^*_\mathrm{c}$, while in the extreme case, in which it is equal to zero, one gets $\mathrm{Ma}^*_\mathrm{c}=24$, which corresponds to completely ignoring the gas phase and just setting a constant solute flux at the interface. Additionally, we noticed that the values of $\sigma$ decrease with increasing values of the factor $c_\mathrm{sat}a_1\rho^\mathrm{g}\VaporDiffusivity_\text{e}$. 

One of the simplifications that we took for the linear stability analysis was to consider the solution as ideal by setting the activity coefficient $\ActivityCoeff_\mathrm{e}=1$. If this is not the case, an increase or decrease of the activity coefficient would affect the value of $c_\mathrm{sat}$ correspondingly. Therefore changes in the activity coefficient would come into effect through $c_\mathrm{sat}$. In particular for ethanol-water mixtures with activity coefficients above one, consideration of the activity coefficient would cause the growth rate $\sigma$ to decrease.

\subsection{Comparison of linear stability analysis with numerical simulations}

\begin{figure}
	\centerline{\includegraphics[width=\linewidth]{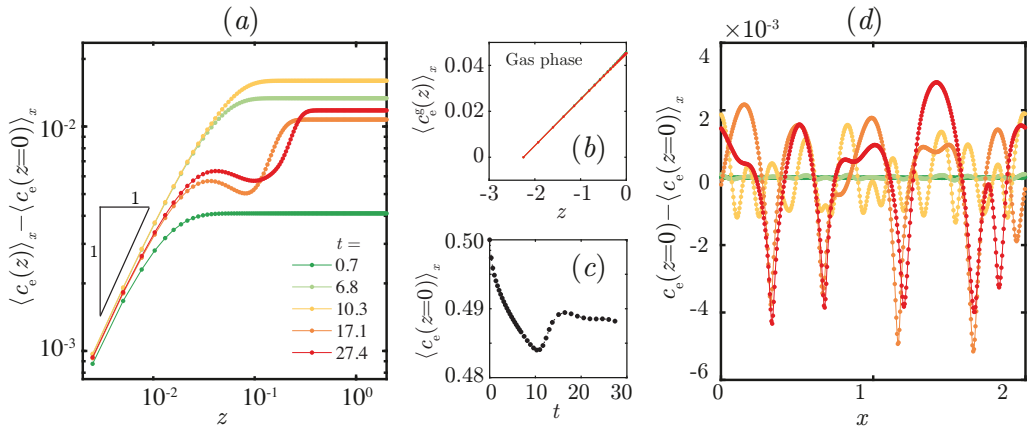}}
	\caption{(\textit{a}) Ethanol mass fraction $\MassFractEth$ averaged over the $x$ coordinate as function of the position $z$ in a log-log axis for a simulation at very early times. The curves were shifted by subtracting the mean mass fraction at $z=0$ to show that close to the edge the profile is actually linear. Different curves represent different times (see legend). (\textit{b}) Mass fraction profile as function of position $z$ on the gas phase for the same times as shown in (\textit{a}). The changes are so small that the curves practically overlap with each other. (\textit{c}) Evolution in time of the averaged mass fraction along the edge $z=0$. (\textit{d}) Mass fraction profile along the edge $z=0$. All curves were shifted by the corresponding averaged mass fraction at the edge. The colors correspond to the same times as those shown in (\textit{a}). Movie 4 of the SI shows the whole concentration and velocity vector fields as function of time.}
	\label{fig:MassFractionEarlyTimes}
\end{figure}

Experimentally, we were not able to observe the onset of the instability since at that moment the accumulation of dye or particles was too low to create a large enough contrast. Furthermore, given that the arches merged with each other, it is possible that by the time we are able to observe them, they are already the result of different merging events, which is what we observe from simulations. Therefore, we relied on the latter for comparison with theory. To do so, we performed simulations over short times, such that merging had not taken place, and in domains at least 5 times larger than the fastest growing wavelength. To test different Marangoni numbers we varied the evaporation rate by changing the size of the gas domain.

In figure \ref{fig:MassFractionEarlyTimes} we show the mass fraction of ethanol $c_\mathrm{e}$ as function of space and time. In particular, these are results for a case of $\MarangoniNumber_\mathrm{s}^* = $ \si{\num{1.55e5}}. In figures \ref{fig:MassFractionEarlyTimes}(\textit{a}) and (\textit{b}) we show the averaged mass fraction profiles on the liquid and gas side, respectively. In both cases the average was taken along the $x$ direction and we display the profiles over 5 different times. In the case of the gas profiles, the changes are so small that they lie on top of each other almost indistinguishably. Notice that we have plotted the liquid profiles on a double logarithmic scale and we have shifted them by subtracting the averaged mass fraction at the interface, i.e. $\langle \MassFractEth(z=0)\rangle_x$ (shown in figure \ref{fig:MassFractionEarlyTimes}(\textit{c})). In this way we can see that close to the interface the profiles are approximately linear, while of course far away they reach the far field value. As in the work by \citet{Kollner_2013_multiscale} we observed a local minimum that results from the mixing zone caused by the Marangoni rolls, indicating that at those times the instability has already developed and that ethanol depleted liquid has already been advected back into the bulk.

\begin{figure}
	\centerline{\includegraphics[width=0.5\linewidth]{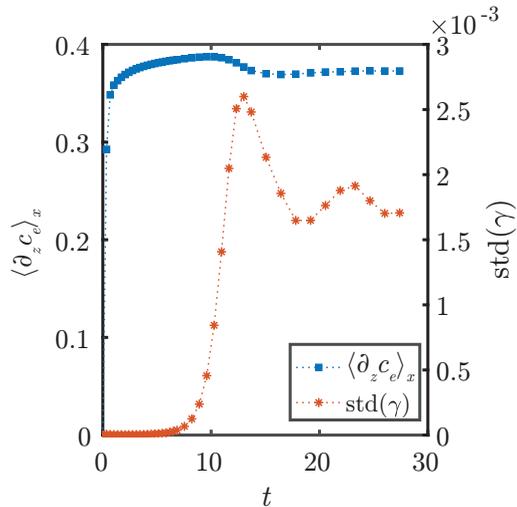}}
	\caption{Average diffusive flux $\langle \partial_z \mathrm{e}\rangle_x$ at the edge and standard deviation of the surface tension $\mathrm{std(}\gamma \mathrm{)}$ as function of time, corresponding to the same simulations as shown in figure \ref{fig:MassFractionEarlyTimes}.}
	\label{fig:Je_stdSurfTen}
\end{figure}

In figure \ref{fig:MassFractionEarlyTimes}(\textit{d}) we show the mass fraction profiles along the edge at different times. These profiles were shifted by their corresponding mean values for visibility. Clearly, the instability has not grown very much after about 7 viscous times, while at $t=10.3$ it has dramatically increased. The presence of the instability is also suggested by the increase of the averaged mass fraction at the edge (figure \ref{fig:MassFractionEarlyTimes}(\textit{c})) and the standard deviation of the surface tension in figure \ref{fig:Je_stdSurfTen}. Furthermore, at $t = 17.1$
merging has already started, showing the very early onset of non-linear effects. 

From the profile at $t = 27.4 $ it can be seen that more than one wavelength is still present at the same time, which can be expected since the perturbation comes from numerical noise combined with the non-uniform shape of the mesh, meaning that more than one mode can be excited at the same time.  Furthermore, certain modes could be excited again, given that at these early times merging takes place by the displacement (along the edge) of the formed rolls, leaving free space for new ones to be generated, as can be seen close to the end of movie 4 of the SI. In fact, we tried cases where we intentionally introduced a perturbation with one particular wavelength, but other wavelengths developed too. Since the results were quite similar, here we present only the cases with random noise. 

The presence of many wavelengths at the same time makes it difficult to determine which one grows the fastest just from looking at the mass fraction profiles, specially since merging starts very soon. In order to answer this question, we calculated the Fourier transform of the edge profiles (figure \ref{fig:MassFractionEarlyTimes}(\textit{d})) at each time. As an example, the profile at $t = 6.8$ is plotted again in frequency space in figure \ref{fig:Power}(\textit{a}) of appendix \ref{app:CalcGrowthRate}. Then we looked at the power as function of time as shown in figure \ref{fig:Power}(\textit{b}). There is a range where the power growth is approximately exponential, and after some time, it saturates and stays at an approximately constant value (at least for this small range of time).

\begin{figure}
	\centerline{\includegraphics[width=0.9\linewidth]{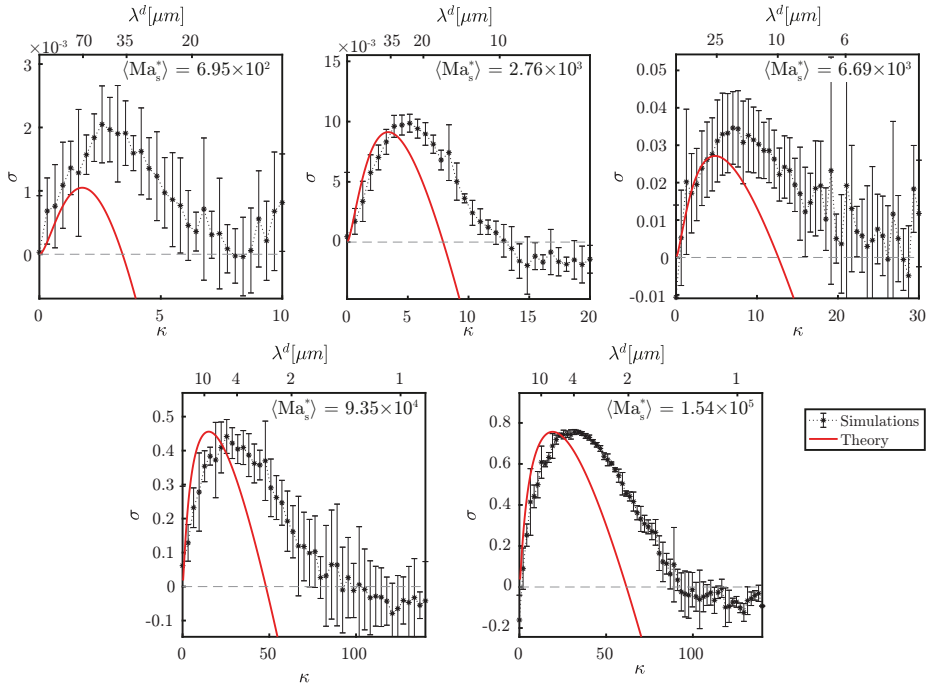}}
	\caption{Comparison of the normalized growth rate $\sigma$ versus normalized wavenumber $\kappa$ obtained from linear instability theory (solid line) and from simulations (*). The upper horizontal axis shows the corresponding wavelength $\lambda^\mathrm{d} = 2\upi\delta^\mathrm{d}/\kappa$. The mean $\langle \MarangoniNumber_\mathrm{s}^*\rangle$ indicated in each plot corresponds to the average value over different realizations (see the SI for a table with the values of $\MarangoniNumber_\mathrm{s}^*$ of individual simulations).}
	\label{fig:Sigma_vs_kappa}
\end{figure}

By fitting a line (in semi-logarithmic scale) to the region shadowed in figure \ref{fig:Power}(\textit{b}), we obtained the growth rate of every single wave number present in the mass fraction profiles. This relies on the assumption that for some time many wavelengths will be present, and not only the fastest one. The corresponding plot of growth rate $\sigma$ versus wave number $\kappa$ is shown in figure \ref{fig:Sigma_vs_kappa} for different $\MarangoniNumber_\mathrm{s}^*$. In each case, the curves are the average over different realizations and the error bars represent one standard deviation on each side. We have included cases using the different kinds of meshes described in section \ref{sec:NS_implementation}. We varied the size of the domain in the $x$ and $z$ directions, but always keeping the domain at least 5 times the size of the fastest growing wavelength in the $x$ direction. We only show cases where we did not impose any initial perturbation and the instability started from numerical error.

Given that in simulations we control the evaporation rate through the size of the gas domain, in order to determine the modified Marangoni number $\MarangoniNumber_\mathrm{s}^*$ of each case, we obtained $E$ from the average diffusive flux at the boundary $\langle \partial_z \MassFractEth \rangle_x$. Figure \ref{fig:Je_stdSurfTen} shows the evolution in time of $\langle \partial_z \MassFractEth \rangle_x$. For the first time steps the mass fraction gradient increases very quickly from zero up to the value determined by the boundary condition (\ref{eq:fluxBCSimplified}), then it increases slowly until merging takes place and the gradient slightly decreases. In all cases, we took the average value over the same period for which we fitted the growth rate of the power (see appendix \ref{app:CalcGrowthRate}). This average mass fraction gradient $ \overline{E} \equiv \overline{\langle \partial_z \MassFractEth \rangle}_x$ was then used to calculate $\MarangoniNumber_\mathrm{s}^*$ together with material quantities corresponding to $\MassFractEth = 0.5$. The corresponding values of $\overline{E}$ and $\MarangoniNumber_\mathrm{s}^*$ obtained for each different run are shown in the SI. The values shown in figure \ref{fig:Sigma_vs_kappa} are the average of the modified Marangoni number over all the different runs. 

With the Marangoni number known, we calculated the theoretical growth rates as function of wavenumber for each case. The results are plotted as solid red lines in figure \ref{fig:Sigma_vs_kappa}. 
The prediction tends to underestimate $\kappa$ by a factor of about 0.5, and for smaller modified Marangoni numbers, the growth rate is underestimated too. Furthermore, we would expect that the inclusion of the activity coefficient would shrink the theoretical curves. However, the agreement is still fairly good considering all the simplifications done for the linear stability analysis, and the fact that in the simulations non-linear terms and water evaporation are taken into account. 

Notice that for large values of $\kappa$, the growth rate calculated from the simulations does not keep on decreasing. This could be a result of the higher resolution needed to better resolve the corresponding wavelengths. Moreover, it can be seen from figure \ref{fig:Power}(\textit{a}) that the power of large values of $\kappa$ ($\kappa > 100$ in that example) is quite small. Therefore, above a given value of $\kappa$ we cannot trust anymore the obtained values of the growth rate.

On top of each of the plots from figure \ref{fig:Sigma_vs_kappa} we have indicated the corresponding values of the dimensional wavelength $\lambda^\mathrm{d} = 2\upi\delta^\mathrm{d}/\kappa$. Note that for the two largest $\MarangoniNumber_\mathrm{s}^*$ the fastest growing wavelength is smaller than the gap between the plates of the cell ($\delta^\mathrm{d} = 20$ \si{\micro\meter}). \citet{Bratsun2004} also found that for large Marangoni numbers their predicted fastest wavelength was smaller than the gap of their cell, which lead them to disregard those solutions. A reason for this is that the Hele-Shaw model relies on the assumption that variations of the velocity in the direction parallel to the plates of the cell are much smaller than variations in the directions normal to the plates \citep{ruyer2001inertial}. Therefore a wavelength smaller than the gap of the cell would result in changes of velocity that would be stronger in the plane of the cell.

In fact from equation (\ref{eqn:MarginalStability}), we can obtain a value for $\MarangoniNumber_\mathrm{s}^*$ for which $\kappa = 2\upi$, meaning that $\lambda^\mathrm{d} = \delta^\mathrm{d}$. This implies that $\MarangoniNumber_\mathrm{s}^* = 8\mathcal{M}(\upi + \sqrt{(\upi^2+3)})^2$, meaning that for Marangoni numbers larger than this quantity, there will be at least one wavelength smaller than the gap. This is already the same regime over which the model becomes close to the purely 2D case (from the point of view of linear stability analysis) suggesting that for large Marangoni numbers a 3D model would be more suitable in order to look at the onset of the instability, even if the gap is as small as 20 \si{\micro\meter}. Nevertheless, as time goes on and coarsening takes place by means of merging, this restriction is not in play anymore, explaining the good agreement between simulations and experiments at much longer times. Furthermore, if the properties of the fluids or the evaporation rate result in a low enough Marangoni number, then equation (\ref{eqn:NonDimDispRel}) could be used as a first estimation of the growth rate of long wavelengths at onset.

\section{Summary, conclusions, and outlook}

We have investigated by means of experiments, numerical simulations, and linear stability analysis the evaporation of a solution of ethanol and water confined in a Hele-Shaw cell. The selective nature of the evaporation process, due to the larger volatility of ethanol, resulted in the appearance of a Marangoni instability at the liquid-air interface. This instability results in the appearance of convective rolls in the bulk of the liquid phase, which caused the accumulation of dye in arch-like regions. With ongoing time, we observed the rolls to grow and merge until they reached the lateral size of the cell. Eventually, the instability stopped and the flow ceased.

Qualitatively, such phenomena were observed before in experiments by \citet{Linde_1964_HS} for different evaporating binary mixtures confined inside a Hele-Shaw cell, however here we measured and analysed the growth of the rolls in a quantitative way by means of confocal microscopy, $\mu$-PIV, and numerical simulations. We found that the size of the arches increases following an overall square-root of time dependence, similarly to
other multicomponent liquid-liquid systems \citep{Kollner_2015}. Furthermore, by looking at averaged velocity profiles, we found that the velocity in the rolls decays exponentially towards the bulk of the liquid.

For numerical simulations and linear stability analysis we considered a quasi-2D model which results from an average over the direction normal to the walls, assuming a parabolic profile for the velocity and a constant mass fraction in that direction. In this way we took into account the drag caused by the walls without having to use the full 3D equations. In spite of these simplifications, simulations and experiments showed a good quantitative agreement supporting the applicability of the quasi-2D model at long enough times.

Due to the merging process that characterizes the evolution of the instability, which also slows down with time, we cannot be sure if the first rolls that we observe experimentally represent the onset of the instability. In fact, from simulations we were able to corroborate that the merging process can start at shorter times than those accessible experimentally. Therefore, for comparison with linear stability theory, we relied only on the results from simulations.

By looking at the perturbation of the mass fraction at the interface at early times, we were able to calculate its growth rate as function of the different wavenumbers present in it.  This allowed us to compare these calculations with the predictions that we made by linear stability theory. The results are in reasonable agreement, taking into account that simulations consider the complete non-linear equations with material properties that depend on mass fraction, while for linear stability analysis we (i) made use of a simplified model with constant material properties, we (ii) disregarded the velocity caused by the evaporation of both ethanol and water, we (iii) did not include the evaporation of water, and we (iv) did not consider a time evolving base state.

The consideration of the drag of the walls of the Hele-Shaw cell in our analysis resulted in the possibility of stability for conditions that would be considered unstable under the purely 2D model of \citet{SternlingScriven1960AIChE}. More specifically, for $\MarangoniNumber_\mathrm{s}^* < \MarangoniNumber^*_c=24\mathcal{M}$ the system is expected to be stable for all values of the wavenumber $\kappa$. For large Marangoni numbers the quasi-2D model tends to the purely 2D model, making it unnecessary to consider the drag of the walls, at least in terms of the linear stability prediction. However, at the same time at large Marangoni numbers, the model predicts wavelengths that can be smaller than the gap of the cell, suggesting that the full 3D equations need to be considered in order to account for the onset of the instability. Nevertheless, as long as the $\MarangoniNumber_\mathrm{s}^* \leq 8\mathcal{M}(\upi + \sqrt{(\upi^2+3)})^2$, the wavelengths predicted by the Hele-Shaw model become smaller than the cell thickness. Therefore, the dispersion relation given by equation (\ref{eqn:NonDimDispRel}) could be used as a first estimate for the initial size of the instability, as long as thermal effects stay small. Moreover, as time goes on and coarsening takes place, the restriction which the gap size imposes is overcome, and, as we mentioned before, the numerical results show good agreement with experiments, even at large values of the Marangoni number. It would be interesting to further test the predictions of the stability analysis using systems with lower solutal Marangoni numbers.

In future work we plan to apply our results to evaporating ternary systems, also confined in a Hele-Shaw cell. In those cases the Marangoni instability will determine the location of spontaneous emulsification (ouzo effect) and could compete with other mechanisms of pattern formation. In particular, this competition could preclude the formation of branch like structures, composed of emulsion droplets, that were observed for a ternary miscible liquid-liquid system \citep{Lu_2017}.
\\

\section*{Acknowledgement} 
Valuable discussion with Yibo Chen, Kai Leong Chong, Luoqin Liu, Yanshen Li, Yaxing Li, Pallav Kant, and Chris Johnson is greatly appreciated. We also thank Alvaro Marin for useful discussion on the micro-PIV technique. 

\section*{Funding}
This work was supported by NWO (ERC-Advanced Grant Project DDD No. 740479, and ERC-Proof-of-Concept Grant Project No. 862032). X.H.Z. acknowledges support from the Natural Science and Engineering Council of Canada (NSERC) and from the Canada Research Chairs program.

\section*{Declaration of Interests} 
The authors report no conflict of interest.

\appendix

\section{Further experimental details} 

\subsection{Pillars and walls inside the chip}\label{app:FurtherExpDetails_Pillars}

The small gap size (\si{\micro\metre}) compared to the lateral dimensions (\si{\centi\metre}) of the Hele-Shaw cell caused a big overhang between the silicon and glass plates. This made the cell prone to collapse at the center during the fabrication process. Therefore we divided the cell into three channels of 1 \si{cm} width each by adding two walls, shown in figure \ref{fig:SketchSetUp}(\textit{b}) as horizontal black stripes. For additional support and to keep a homogeneous gap, a matrix of pillars, 50 \si{\micro\metre} in diameter, was added in between the walls. In figure \ref{fig:SketchSetUp}(\textit{a}), the pillars are represented as thin vertical lines, while in figure \ref{fig:SketchSetUp}(\textit{b}) the gap region is colored in gray, and the pillars are shown as black circles (their size is exaggerated).

\subsection{Surface coating}\label{app:FurtherExpDetails_Coating}
The surface of the chip was made hydrophobic by depositing a layer of Trichloro (octadecyl) silane (OTS, Sigma-Aldrich) by chemical liquid deposition in a similar way as in \cite{Zhang2006_Langmuir_Nanobubbles}. To activate the surface, we dipped the chip in a bath of piranha solution (30\% of hydrogen peroxide and 70\% sulfuric acid) for 40 \si{\minute} at 75\si{\degreeCelsius{}}. We then washed it with plenty of Milli-Q water and ethanol. The chip was kept in an oven at 120\si{\degreeCelsius } for 2 \si{\hour} to eliminate traces of water. After cooling down, the chip was dipped in a solution of 0.5\% OTS in toluene for 17 \si{\hour}. Finally the chip was sonicated for 10  \si{\minute}  in chloroform and 15  \si{\minute}  in each of the following solvents one after the other: toluene, acetone, isopropyl alcohol, ethanol and water. Finally, we dried the chip under a nitrogen flow.

\subsection{$\mu$-PIV}\label{app:Experimental details micro-PIV}

A suspension of fluorescent particles (FluoSpheres, Life Technologies, 1 \si{\micro\meter}, carboxilate-modified, 2\% solid in aqueous solution, with 2 mM of sodium azide) was added at 0.9 wt\% to the ethanol-water solution. This resulted in an initial volume fraction of particles of 0.016 \% and mass fraction of sodium azide of 1.2$\times 10^{-4}$ wt\%, meaning that we can disregard any effects from the latter even after 30 min of evaporation.

We recorded images with a confocal microscope (Nikon Confocal Microscopes A1 system) at 30 fps using a 488 \si{\nano\meter} laser and a pinhole size of 26.8 \si{\micro\meter}. This settings resulted in a vertical resolution of 8 \si{\micro\meter}. We made measurements at a single position in the $y$ direction, approximately in the middle of the cell. By translating the position of the focal plane along the gap of the cell, we observed that the visible particles were always the same, but with different intensities. Therefore, despite the vertical resolution of 8 \si{\micro\meter} all the particles inside the gap were visible at the same time. Different particles moved at different speeds, which we attribute to the Poiseuille flow. Therefore, the velocity results shown in this work represent a weighted average of the velocity, where the dominant peak during the correlation step in the PIV process determines the velocity at a given position.
 
The recorded images were processed with PIVLab for MatLab \citep{thielicke2014pivlab, Thielicke2019, Thielicke2014PhD}. We used interrogation windows of 40 pixels $\times$ 40 pixels (100 \si{\micro\meter} $\times$ 100 \si{\micro\meter}) with a 50\% overlap.

We calculated the relaxation time according to \citep{Tan_2019, Li_2019_Bouncing}
\begin{equation}
    t_0^\mathrm{d}\equiv \left(1+\frac{\rho_\mathrm{sol}^\mathrm{d}}{2\rho_\mathrm{p}^\mathrm{d}}\right)\frac{(d_\mathrm{p}^\mathrm{d})^2\rho_\mathrm{p}^\mathrm{d}}{18\mu_\mathrm{sol}^\mathrm{d}},
\end{equation}
where $\rho_\mathrm{sol}^\mathrm{d} \approx 910.73$ \si{\kilo\gram/\meter^3} and $\mu_\mathrm{sol}^\mathrm{d}\approx 2.5$ \si{\milli\pascal\second} are the density and dynamic viscosity of the ethanol-water solution, and $\rho_\mathrm{p}^\mathrm{d} = 1050.00$ \si{\kilo\gram/\meter^3} and $d_\mathrm{p}^\mathrm{d} = 1$  \si{\micro\meter} are the density and diameter of the tracer particles. Then the Stokes number is estimated to be $\mathrm{St} = t_0^\mathrm{d}u^\mathrm{d}_\mathrm{max}/z_\mathrm{F}^\text{d}$, with $u^\mathrm{d}_\mathrm{max}$ the maximum velocity at a given time, and $z_\mathrm{F}^\text{d}$ the roll height at that time. The velocity ranges from $\sim 10^{-4} - 10^{-3}$ m/s while the roll height goes from $0.1 - 1$ \si{\milli\meter}, resulting in a Stokes number of at most $\mathrm{St}\sim 10
^{-7}$. If we consider that from simulations we observed that the velocity at the edge reaches up to 0.01 m/s and we combine this with the smallest roll that we can observe experimentally, then we get $\mathrm{St}\sim10^{-6}$, meaning that in any case the particles should follow the flow properly.

\section{Values of different properties used for simulations and theoretical calculations}\label{app:TableOfValues}

In table \ref{tab:DimensionalValues} we show the initial dimensional values of the different properties of the fluids used in the simulations. As mentioned in section \ref{sec:ModelEqns}, the generalized form of Rault's law (equation \ref{eqn:RaoultsLaw}) was used in simulations, for which the activity coefficient was calculated by the thermodynamic model AIOMFAC \citep{Zuend_2008, Zuend_2011}. On the other side, as also mentioned before, a Taylor expansion around an initial mass fraction of $c_\mathrm{e,0} = 0.5$ was used for the theoretical calculations. The corresponding coefficients were calculated according to $a_0
(c_\mathrm{e,0} = 0.5)= (1+M_\mathrm{e}^\mathrm{d}/M_\mathrm{w}^\mathrm{d})^{-1}-2(M_\mathrm{e}^\mathrm{d}/M_\mathrm{w}^\mathrm{d})(1+M_\mathrm{e}^\mathrm{d}/M_\mathrm{w}^\mathrm{d})^{-2}$ and $a_1(c_\mathrm{e,0} = 0.5) = 4(M_\mathrm{e}^\mathrm{d}/M_\mathrm{w}^\mathrm{d})(1+M_\mathrm{e}^\mathrm{d}/M_\mathrm{w}^\mathrm{d})^{-2}$. Their numerical values are displayed in table \ref{tab:DimensionalValues}. 

The thermal diffusion coefficient was obtained from \citet{yano1988thermal}, corresponding to $\MassFractEth=0.494$ at 288 K. The gradient of surface tension with respect to temperature was obtained by fitting data from \citet{vazquez1995surface}. 

For the case of $E$, as discussed in the main text, we did not used the value at $t=0$ as there is no gradient initially. Instead we used the mean value $\overline{E}$ calculated over the same period used for calculating the growth rate $\sigma$ as described in section \ref{app:CalcGrowthRate}. The corresponding values for each simulation can be seen in the SI.

\begin{table}
	\begin{center}
		\def~{\hphantom{0}}
		\begin{tabular}{llr|llr}
    		$c_\mathrm{e,0}$ &  & 0.5  & $c_\mathrm{e,0}^\mathrm{g}$ &  & 0 \\
			$D^\text{d}_0$ &[\si{\meter\squared\per\second}] & \si{\num{3.65e-10}} & $D^\text{g,d}_{e,0}$ &[\si{\meter\squared\per\second}] & \si{\num{1.35e-5}} \\
			$\rho^\text{d}_0$ &[\si{\kilogram\per\meter\cubed}] & \si{\num{9.11e2}}  & $\rho^\text{g,d}_0$ &[\si{\kilogram\per\meter\cubed}] & \si{\num{1.2}} \\
			$\mu^\text{d}_0$ &[\si{\kilogram\per\meter\per\second}] & \si{\num{2.49e-3}} & $\mu^\text{g,d}_0$ &[\si{\kilogram\per\meter\per\second}] & \si{\num{1.80e-5}}\\ $\frac{d\gamma^\text{d}_0}{dc_\mathrm{e}}$ &[\si{\kilogram\per\second\squared}] & \si{\num{-1.85e-2}} & $D^\text{g,d}_{w,0}$ &[\si{\meter\squared\per\second}] & \si{\num{2.45e-5}} \\
			$p_\mathrm{sat}^\text{d}$ &[\si{\kilogram\per\meter\per\second\squared}] & \si{\num{6.57e3}} & $T^\mathrm{d}$ & [\si{\K}]& \si{\num{2.95e2}}\\ 
			$M_\mathrm{e}^\mathrm{d}$ &[\si{\kilogram\per\mol}] & \si{\num{4.61e-2}} & $M_\mathrm{w}^\mathrm{d}$ &[\si{\kilogram\per\mol}] & \si{\num{1.80e-2}}\\ 
			RH & [\%] & 50  & $\delta^\mathrm{d}$ &[\si{\meter}] & \si{\num{2.00e-5}}  \\
			$a_0$ &  & \si{\num{-1.23e-1}}  & $a_1$ & & \si{\num{8.08e-1}} \\
			$D^\text{d}_\mathrm{0,T}$ &[\si{\meter\squared\per\second}] & \si{\num{8.74e-8}} & $\frac{d\gamma^\text{d}_0}{dT}$ &[\si{\kilogram\per\second\squared}] & \si{\num{-8.99e-5}} \\
		\end{tabular}
		\caption{Dimensional values used in simulations for the two phases at $t=0$ and for the thermal Marangoni number estimation.}
		\label{tab:DimensionalValues}
	\end{center}
\end{table}

\section{Derivation of dispersion relation }\label{app:DerivationDispRelation}

Here we show in more detail the derivation of the dispersion relation (\ref{eqn:NonDimDispRel}). 
Introduction of the perturbed quantities (\ref{eqn:Pert_u}) to (\ref{eqn:Pert_cg}) into equations (\ref{eq:LS:contitwodim}) to (\ref{eq:LS:Laplacetwod}) yields the following ordinary differential equations after neglecting nonlinear terms on $\epsilon$,

\begin{eqnarray}
& &\left(\frac{d^2}{dz^2} - \kappa^2\right)\hat{p}_1 = 0, \label{eqn:ODEp1}\\
& &\left(\frac{d^2}{dz^2} - \kappa^2 - 12 - \sigma  \right)\pmb{\hat{u}}_1 =  \left(\frac{d}{dz}\hat{z} + i\kappa\hat{x}  \right)\hat{p}_1,\label{eqn:ODEu1}\\
& &\left( \frac{d^2 }{d z^2} -\kappa^2 - \sigma\SchmidtNumber\right) \hat{c}_\mathrm{e,1}(z) = E\SchmidtNumber\hat{u}_{z,1}(z) \label{eqn:ODEc1}\\
& &\left( \frac{d^2 }{d z^2} -\kappa^2 \right) \hat{c}_\mathrm{e,1}^\mathrm{g}(z) = 0, \label{eqn:ODEc1_g}
\end{eqnarray}

Equation (\ref{eqn:ODEp1}) was obtained in the regular way of applying the divergence to the Navier-Stokes equation in combination with the continuity equation. It has for solution 
\begin{equation} \label{eqn:App_p1}
\hat{p}_1 = f_p e^{-\kappa z},
\end{equation}
where $f_p$ is a constant and we have kept $\hat{p}_1$ finite.
Substituting equation (\ref{eqn:App_p1}) into equation (\ref{eqn:ODEu1}) leads to the solutions
\begin{equation} \label{eqn:App_w1}
\hat{u}_{z,1} = -\frac{\kappa f_p}{\kappa^2-\beta^2}\left(e^{- \kappa z} - e^{-\beta z}\right),
\end{equation}
in the $z$ direction, and 
\begin{equation} \label{eqn:App_u1}
\hat{u}_{x,1} =\frac{if_p}{\kappa^2-\beta^2}\left(\kappa e^{- \kappa z} - \beta e^{-\beta z}\right)
\end{equation}
in the $x$ direction, with $\beta^2 = \kappa^2 + 12 + \sigma$. In these solutions we have already considered that the velocity should stay finite, assumed $u_{z,1}(x,0,t)=0$ as described in the main text, and took into account continuity $\pmb{\nabla}_{\parallel}\cdot\pmb{u} = 0$.

Introduction of equation (\ref{eqn:App_w1}) into equation (\ref{eqn:ODEc1}) leads to

\begin{equation} \label{eqn:App_c1}
\hat{c}_\mathrm{e,1} = \frac{B_0}{\kappa^2-\zeta^2} e^{- \kappa z} - \frac{B_0}{\beta^2-\zeta^2}e^{-\beta z} + B_1 e^{-\zeta z},
\end{equation}
where $\zeta^2 = \kappa^2 + \sigma\SchmidtNumber$, $B_0 = - E \SchmidtNumber f_p \kappa/(\kappa^2-\beta^2)$, and $B_1$ is a constant. Again we took a finite perturbation far away from the edge.

The solution of equation (\ref{eqn:ODEc1_g}) is given by
\begin{equation} \label{eqn:App_c1_g}
\hat{c}_\mathrm{e,1}^\mathrm{g} = B_1^\mathrm{g} e^{\kappa z},
\end{equation}
with $B_1^\mathrm{g}$ a constant. Once more, we have consider that the perturbation tends to zero far away from the edge.

The expanded Raoult's law (\ref{eq:Raoults_Expanded}) for the base case results in
\begin{equation}
\mathcal{C}^\mathrm{g}_0  = c_\mathrm{sat}(a_0 + a_1 \mathcal{C}_0),
\end{equation}
meaning that for the perturbed mass fraction the condition reads
\begin{equation}
\hat{c}_\mathrm{e,1}^\mathrm{g}(0)  = c_\mathrm{sat}a_1 \hat{c}_\mathrm{e,1}(0),
\end{equation}

thus
\begin{equation}
	B_1^\mathrm{g}= c_\mathrm{sat} a_1 (I(0)+B_1),
\end{equation}
where $I(0) = \frac{B_0}{p}  - \frac{B_0}{q}$, with $p = \kappa^2 - \zeta^2$ and  $q = \beta^2-\zeta^2$.

On the other side, the mass conservation of solute at the interface (\ref{eq:fluxBCSimplified}) for the base case results in 
\begin{equation}
 E =  D^\mathrm{g}_\mathrm{e}\rho^\mathrm{g}E^\mathrm{g},
\end{equation}
 while for the perturbed mass fraction we have 
\begin{equation} \label{eqn:BCFluxPerturb}
I'(0) + \zeta B_1 = - D^\mathrm{g}_\mathrm{e}\rho^\mathrm{g} \kappa B_1^\mathrm{g},
\end{equation}
with $I^{\prime}(0) = \frac{\kappa B_0}{p}  - \frac{\beta B_0}{q}$, and $\VaporDiffusivity_\mathrm{e} = D^\mathrm{g,d}_\mathrm{e}/D_0^\mathrm{d}$.

In this way we can obtain $B_1$ and $B_1^\mathrm{g}$ in terms of $B_0$, but more importantly the mass fraction at the edge, which in the liquid side reads
\begin{equation} 
\hat{c}_\mathrm{e,1}(0) = \frac{\zeta I(0) - I^{\prime}(0)}{\kappa c_\mathrm{sat}a_1 D^\mathrm{g}\rho^\mathrm{g}  + \zeta }.
\end{equation}

As a last step we apply boundary condition (\ref{eq:SurfTenBC_LS}),
\begin{equation} 
\frac{d \hat{u}_{x,1}(0)}{dz} =i\kappa \frac{Ma}{\SchmidtNumber} \hat{c}_\mathrm{e,1}(0),
\end{equation}
where again we have taken into account that $u_{z,1}(x,0,t)=0$. In this way, the dispersion relation is expressed as 
\begin{equation} \label{eqn:AppDispRel}
 E\MarangoniNumber_\mathrm{s} = \kappa^2 \left( c_\mathrm{sat}a_1 D^\mathrm{g}\rho^\mathrm{g} + \frac{\zeta}{\kappa}\right)\left(1+\frac{\beta}{\kappa}\right)\left(1 +\frac{\zeta}{\kappa}\right)\left(\frac{\beta}{\kappa}  + \frac{\zeta}{\kappa}\right) ,
\end{equation}
 which becomes equivalent to equation (30) from \citet{SternlingScriven1960AIChE}, provided that we neglect surface viscosity, and consider that $\mu^\mathrm{g,d}\ll\mu^\mathrm{l,d}$, $\rho^\mathrm{g,d} \ll \rho^\mathrm{l,d}$, and $D^\mathrm{l,d}\ll D^\mathrm{g,d}$, with the superscripts $\mathrm{g}$ and $\mathrm{l}$ referring to the gas and liquid phases, respectively. We solved the complete equation from \citet{SternlingScriven1960AIChE} for a few cases and compared it with its simplified counterpart (\ref{eqn:AppDispRel}) resulting in negligible differences when taking into account the material properties of ethanol and water and keeping the surface viscosity term equal to zero.

Finally, if we notice that $\MarangoniNumber_\mathrm{s}^* = E \MarangoniNumber_\mathrm{s}$, then equation (\ref{eqn:AppDispRel}) leads to equation (\ref{eqn:NonDimDispRel}) in the main text.

\section{Calculation of the growth rate }\label{app:CalcGrowthRate}

\begin{figure}
	\centerline{\includegraphics[width=\linewidth]{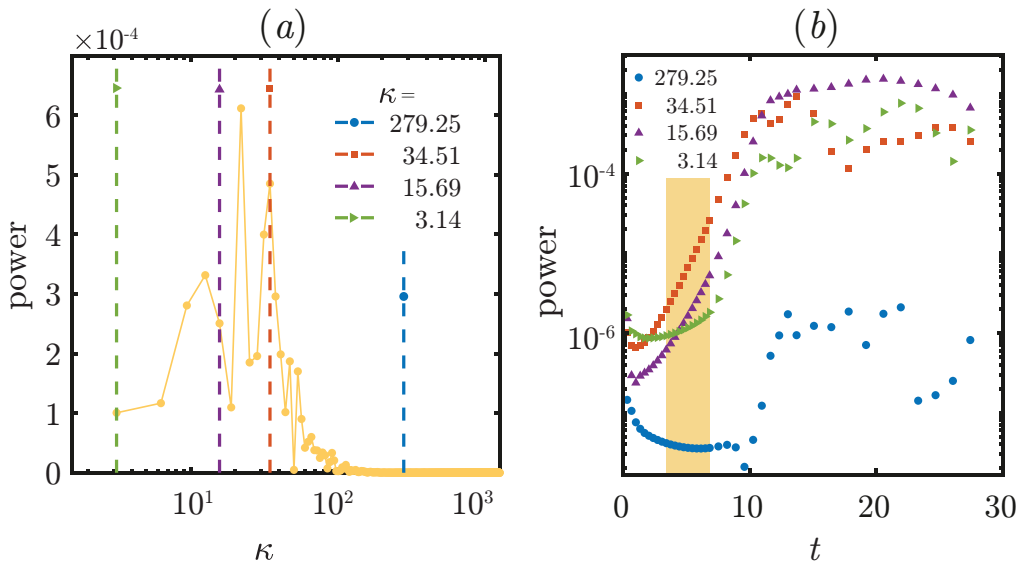}}
	\caption{(\textit{a}) Power density function obtained from the fast Fourier transform of the mass fraction profile at $t = 10.3$ from figure \ref{fig:MassFractionEarlyTimes}(\textit{d}) as function of the normalized wavenumber $\kappa$. (\textit{b}) Same power as in (\textit{a}) but now as function of time in a semi-log scale. The four curves correspond to the vertical dashed lines in (\textit{a}). The shadowed area represents the section over which a linear fit (in semi-log scale) was used to calculate the growth rate $\sigma$.}
	\label{fig:Power}
\end{figure}

Here we show typical plots used for the calculation of the growth rate $\sigma$ as function of wavenumber $\kappa$. In figure \ref{fig:Power}(\textit{a}) we show the power density function (PDF) as function of the normalized wavenumber $\kappa$. This PDF corresponds to the mass fraction profile at $t = 10.3$ from figure \ref{fig:MassFractionEarlyTimes}(\textit{d}). By calculating the PDF over different times, we can look at the evolution in time of the power for every single wavenumber (not only those for which the power has a peak). In figure  \ref{fig:Power}(\textit{b}) we show four examples corresponding to the vertical dashed lines in figure \ref{fig:Power}(\textit{a}). 

If plotted in a semilog scale, there is a region where the power can be approximately fitted by a straight line (shadowed region in figure \ref{fig:Power}(\textit{b})), reflecting exponential growth. The growth rate was obtained from a straight line fit in the shadowed region. We did not apply the fit for earlier times to avoid considering the period over which the concentration gradient is building up (transients, see figure \ref{fig:Je_stdSurfTen}), but also not at later times, because eventually either non-linear coupling or spectral bleeding causes all the wavelengths to grow at the same rate as the fastest wavelength. One can see that both the blue line with circles and the green line with right triangles in figure \ref{fig:Power}(\textit{b}) eventually start growing at a faster rate before reaching a saturation value. However, in some cases we included part of these regions of non-linear growth (for the largest wave numbers), in order to have enough fitting points for the smallest wave numbers, which is the reason for the noisy behavior observed in figure \ref{fig:Sigma_vs_kappa} at larger values of $\kappa$. 

Figure \ref{fig:Sigma_vs_kappa} shows the average over different runs, where we have changed the mesh kind (see SI for a description of the different kinds of meshes), the resolution and the size of the simulation domain. The use of different resolutions and different domain sizes, caused that the specific values of $\kappa$ obtained for the Fourier transform to be different between simulations. In order to obtain an average curve of $\sigma(\kappa)$, we interpolated the values of $\sigma$ over a set of values of $\kappa$ common to every different run. 

From figure \ref{fig:Power}(\textit{b}) it is evident that the power is quite small in the region over which we made the fit (notice that the power represents variations on the mass fraction). However, by the time the value becomes large, either merging has already taken place or the fastest wavelength has already shadowed the growth rate of the others. Therefore, we made the fit at early times when different wavelengths have different growth rates. In movie 5 of the SI we show the evolution in time of the mass fraction variation showing that merging already takes place before the variations in mass fraction becomes large. This early merging makes it hard to observe the onset of the instability in experiments.

In some of our simulations the instability was triggered first in certain regions of the interface, while it started later in others. We discarded these simulations. 

\bibliographystyle{jfm}
\bibliography{references}

\end{document}